%% file: ms.tex
   \documentclass[12pt,preprint]{aastex} 

\newcommand{\Ni}{\rm ^{56}Ni}
\newcommand{\co}{\rm ^{56}Co}
\newcommand{\fe}{\rm ^{56}Fe}
\newcommand{\nni}{{\rm Ni}}

\newcommand{\seco}{\rm~s}
\newcommand{\Msun}{~M_\odot}
\newcommand{\msun}{M_\odot}
\newcommand{\gcm}{\rm ~g~cm^{-3}}
\newcommand{\cmsg}{\rm ~cm^{2}~g^{-1}}

\newcommand{\cmc}{\rm ~cm^{-3}}

\newcommand{\kms}{\rm ~km~s^{-1}}
\newcommand{\cms}{\rm ~cm~s^{-1}}
\newcommand{\erg}{\rm ~ergs}

\newcommand{\ergcc}{\rm ~erg~cm^{-3}}

\newcommand{\vf}{v_{f}}

\newcommand{\Ms}{M_{s}}

\begin{document}


\title {
  NICKEL BUBBLE EXPANSION IN TYPE Ia SUPERNOVAE: ADIABATIC SOLUTIONS \\
  {\it Astrophysical Journal, 2008, Accepted}
 }

\author{Chih-Yueh Wang}
\affil{Department of Physics, Chung-Yuan Christian University \\
22 Pu-Jen, Pu-Chung Li, Chung-Li 32023, Taiwan}
  \email{yueh@phys.cycu.edu.tw}

\begin{abstract}


  This study investigates the expansion properties of nickel (Ni) bubbles
  in Type Ia supernovae (SNe Ia) 
  due to radioactive heating from the $\Ni \rightarrow\ \co \rightarrow\ \fe$ decay sequence,
  under the spherically symmetric approximation. 
    An exponentially-declining medium is considered as the ejecta substrate,
   allowing for the density gradient expected in a Type Ia supernova (SN Ia).
 The heating gives rise to an inflated Ni bubble, which induces 
 a forward shock that compresses the outer ambient gas into a shell.
  As the heating saturates, 
    the flow tends toward a freely-expanding state with the structure frozen into the ejecta.
  The thickness of the shell is $\sim 1\%$ of the radius of the bubble,
  and the density contrast across the shell reaches $\chi \ga 100$ in a narrow region that is limited by numerical
resolution.

  In the adiabatic case (in which no radiative energy diffuses across the shell)  
 the structure of the shell can be approximately described by a self-similar solution  
that is determined by the expansion rate and ambient density gradient of the shell.
 In the radiative case, the shell expansion weakens but remains comparable to 
 the adiabatic solution.
     The density contrast of the inferred ejecta clumps (shell components) is greater than
     that given by the model that uses a uniform ejecta substrate,   
       while the interaction of the clumps with the remnant 
       is delayed to a more advanced evolutionary stage. 
   The explosion parameters of the SN are varied to examine whether the created clump characteristics 
   are consistent with those of the ejecta knots that are present at the edges of Tycho's remnant. 
   Explosion conditions similar to those of successful explosion models are found,
    including the deflagration W7 and the delayed detonation
   yield with sufficient ejection velocities as well as timely initiations for the clump-remnant interaction,
  whereas 
   the luminous helium (He) detonations   
 and the low-energy Chandrasekhar-mass explosions are unfavorable.
    The clump speed can be increased 
     as the initial density contrast of $\Ni$, $0.5 \la \omega < 1$, is reduced,  
   as determined by a realistic elemental distribution.
In all cases, the occurrence of the reverse shock (RS) impact of the clump with the remnant is 
expected in under 2000 yrs after the supernova explodes.

\end{abstract}

 \keywords{hydrodynamics -- nuclear reactions, nucleosynthesis, abundances ---  
            supernova remnants --- supernovae: general }

\section{INTRODUCTION}

Observations of SN 1987A have shown that the distribution of Fe in the ejecta is not what
would be expected in the simplest models; it has higher velocities
than expected and a large filling factor for its mass of
$0.07\Msun$, as determined from the supernova light curve (McCray 1993; Li {\it et al.} 1993).
A plausible mechanism for the large filling factor is the Ni bubble
effect, in which the radioactive progenitors of the Fe expand
relative to their surroundings because of the deposition of radioactive
power (Woosley 1988; Li {\it et al.} 1993; Basko 1994).
This effect is important during the first $\sim$ ten days after the supernova is formed,
when the radioactive power is significant and the
 diffusion of energy has not yet become important.

An expected effect of the Ni bubble expansion is to create clumps
in the nonradioactive gas of intermediate elements.
Widespread evidence indicates the ejecta of core-collapse supernovae are clumpy.
The oxygen line profiles in the nearby Type II supernovae SN 1987A
(Stathakis {\it et al.} 1991) and SN 1993J (Spyromilio 1994; Matheson {\it et al.} 2000) showed
evidence of the structure, implying that the gas is clumped.
The velocity range of the emission extends to 
$1,500\kms$ in SN 1987A and $4,000\kms$ in SN 1993J.
 Clumping was also evident in 
  the prototypes of the oxygen-rich supernova remnants (SNRs) Cassiopeia A (Cas A) (Hughes {\it et al.} 2000; Hwang {\it et al.} 2000) and 
  Puppis A (Winkler {\it et al.} 1988), 
 and in the Type Ib SN 1985F (Filippenko \& Sargent 1989). 
 As for SNe Ia, 
          observations of SN1006 suggested that 
  Si freely expands in the velocity range 5,600 - 7,000 $\kms$, 
  while its Fe mass is only estimated as $0.075-0.16 \Msun$ (Hamilton {\it et al.} 1997).
 Moreover, 
 in Tycho's SNR (SN 1572), 
 X-ray observations revealed two knots, 
 one enriched in Si and the other in Fe, 
  protruding non-deceleratingly from the SE edge of the remnant 
 at a velocity of $8,300 (D/2.5\ \rm kpc) \kms$, 
 where $D$ is the distance (Vancura {\it et al.} 1995; Hwang \& Gotthelf 1997; Hughes 1997).  
 X-ray spectra of Tycho's remnant 
further suggested that
the Fe line emitting gas is in general at a higher temperature and lower density than 
than the Si line emitting gas (Hwang {\it et al.} 1998); 
this result is also expected from the Ni bubble effect.

  Wang \& Chevalier (2001 and 2002, hereafter WC01 and WC02) 
  studied the role of ejecta clumps in the evolution of SNRs.
The ejecta knots associated with Tycho's remnant
require a free expansion velocity $v \sim 7,000 \kms$ and 
a density contrast $\chi\ga 100$ relative to the surrounding ejecta,
to survive crushing and remain distinct features in the remnant in the present epoch (WC01).
The remarkable protrusions (Aschenbach {\it et al.} 1995) 
on the periphery of the Vela SNR  
     are also likely to be caused by ejecta clumps, 
   and we estimated that $\chi \sim 1000$ and $v \sim 3,000$ are needed to create the structure (WC02).
     My previous work (Wang 2005, hereafter W05)   
        further examined 
    the process of $\Ni$ radioactive heating 
   as a plausible mechanism for forming such ejecta clumps, 
     taking into account the effect of radiative diffusion. 
   In the case of SNe Ia,
  the created clump properties seem compatible with those needed by Tycho's knots.
   This result nevertheless contrasts with the case of core-collapse SNe like Vela's bullets,
   whose required high compression is not anticipated in the simplest Ni bubble scenario.

 Evidence may exist that for SNe Ia, ejecta clumping occurs around one large central Ni bubble. 
 In radio observations of Tycho's remnant, Reynoso {\it et al.} (1997) and Velazquez {\it et al.} (1998) 
 found a wide and regularly-spaced emission in the NW edge of the remnant,
which extends very close to the blast wave.
 The large radius of the structure is not attributable to 
   the intershock instabilities that grow from small perturbations, 
but is effectively explained by the expansion of a spherical shell of clumpy ejecta  
into the intershock region of the remnant (WC01).
In fact,
compared with various extensively observed Galactic Type II SNRs  
which in general exhibit quite complex abundance structures
 	 (such as in Hughes {\it et al.} 2000 and Hwang, {\it et al.} 2000 for Cas A; 
        and Park {\it et al.} 2002 for G292.0+1.8),
both observations 
and modeling of X-ray emissions of Type Ia SNRs
revealed a simple and stratified composition 
(Badenes {\it et al.} 2003, 2005; Decourchelle {\it et al.} 2001 for Tycho; and Lewis {\it et al.} 2003 for N103B).
The stratification of ejecta of SNe Ia seems to be a natural consequence of the lack of neutrino-driven convection (Kifonidis {\it et al.} 2000)
and pulsar winds (Blondin {\it et al.} 2001b) in them,
                which two mechanisms can mix elements and enhance compression.   
      Conversely, the disturbance that is expected in core-collapse SNe
         may greatly complicate the evolution of the Ni bubble on a large scale.

   Dwarkadas \& Chevalier (1998, hereafter DC98)
     	   reviewed the solutions 
  	   for the density structure of SNe Ia
obtained by various 1-D explosion models; 
 they found that an exponential profile, 
which effectively evolves from a steep power law profile to a flatter one,
is the best overall approximation to the density profile  
that is obtained by summing over individual chemical elements. 
            However, in W05, 
            a uniform density distribution 
  (comparable to the inner flat component of an $n=8$ power-law ejecta density model with the density
	      $\rho \propto r^{-n}$ in the outer parts) was assumed,
  	     so the results may not be suitable for SNe Ia.
              Furthermore, 
               the W05 models of SNe Ia show that 
      the propagation of the Ni bubble is only slightly faster than the free motion of the ejecta, 
   suggesting that the case of SNe Ia is more susceptible than the case of core-collapse SNe
   to the ejecta substrate structure.
  However, how the actual exponential decline in density affects the solution is unclear.

 	  The aim of this work is to determine the properties of the Ni bubble-driven shell 
             as ejecta clumps
	  associated with SNe Ia, using a realistic ejecta structure. 
	    A simple, adiabatic and spherically symmetric scenario is first considered to determine
how an evolving exponential density gradient 
influences the predicted clump properties. 
             Radiative-transport radiation hydrodynamics (RHD) are then adopted to include
   the effect of the diffusion of radiative energy across the Ni bubble structure  
  to show how the radiative solution differs. 
	  The explosion parameters used in our calculations are similar to those of the 1-D explosion models of
          Hoflich \& Khokhlov (1996, hereafter HK96) and Hoflich {\it et al.} (1998, hereafter HWK98).
	   We note that 
 	   the radiative transport process due the $\Ni$ radioactive decay 
	   has not before been incorporated
	   in the hydrodynamic (HD) phase of the explosion models,
              and 
 such a proper treatment is particularly difficult for numerical convergence in the case of SNe Ia .
	  \S~2 elucidates our computational setup and methods.
	   \S~3 presents the evolutionary properties of the $\nni$ bubble shell.
	  \S~4 draws on the self-similar solution for a pulsar bubble,
	    to provide insight into the shell structure.
	  \S~5 
          compares the inferred ejecta clump properties 
          of various exponential models.
	 \S~6 draws conclusions.

	\section {DENSITY STRUCTURE IN TYPE Ia SUPERNOVAE}

	SNe Ia are widely regarded as being  
         thermonuclear explosions of carbon-oxygen (C-O) white dwarfs (WDs). 
	Soon after the explosion, the ejecta freely expand
	so that each gas element moves with a constant velocity $v=r/t$ 
	and its density drops to $t^{-3}$, as in a spherical expansion.
	 DC98 showed that, in this phase, the density distribution 
         of valid SN Ia explosion models
	 can generally be described by an exponential profile, 
	\begin{equation}
	\rho_{SN} = A \exp(-v/v_e) \  t^{-3},
	\end{equation}
	  where $A$ is a constant and $v_e$ is another constant called the velocity scale height, 
        which are determined by
	 the total mass $M$ and kinetic energy $E$ of the ejecta.
	  The two constants can be obtained by integrating the mass density and the kinetic energy density over space: 
	$M = 8 \pi A v_e^3$ and $E=48 \pi A v_e^5$, 
	or 
	\begin{equation}
	 v_e = (E/6M)^{1/2} = 2.44 \times 10^8
		\   {M_{1.4}^{-1/2}} \  {E_{51}^{1/2}}  \ \  \rm{cm} \ \rm{s^{-1}},
	\end{equation}
	\begin{equation}
	 A = {6^{3/2}\over 8\pi} {M^{5/2}\over E^{3/2}} =
	     7.67 \times 10^6  \ {M_{1.4}^{5/2}} \  E_{51}^{- {3/2}}
	 \ \ \rm{g} \ \rm{s^3}\  \rm{cm^{-3}},
	\end{equation}
 	where $M_{1.4}$ is the explosion mass in terms of the Chandrasekhar mass, $1.4 \msun$,  
	 and $E_{51}$ is the explosion energy in units of $10^{51} \erg$.
	 A comparison with the power-law model $\rho \propto v^{-n}t^{-3}$ (Chevalier \& Liang, 1989)
          yield
 	 the approximate power index of the density, 
	  $n =  - d ln \rho/ d ln r =  v/v_{e} = r/v_{e}t$.
	 Hence, 
	 models with higher velocity scale (increasing with the $E/M$ ratio) 
	 yield a flatter density distribution at a given flow velocity, 
	 and the gradient of the density increases with radius and decreases over time.

	The creation of an exponential profile for SNe Ia
	is probably related to 
        the fact that the explosion energy is steadily released behind a burning front,
        unlike in the case of pure shock acceleration in core-collapse SNe.
	Three progenitor scenarios have been proposed to distinguish SNe Ia
	(HK96; HWK98; for summary see Hoflich {\it et al.} 2003).
The most favored is the explosion of a near-Chandrasekhar mass C-O WD 
 that has accreted mass through Roche-lobe overflow from an evolved companion star, 
 triggered by compressional heating near the WD center.
In this scenario, 
ignition is initiated in the central high-density region of the WD.
 In the deflagration model, 
the burning front propagates subsonically through the white dwarf; 
  a popular example is the W7 model of Nomoto {\it et al.} (1984). 
In the other class of models, which includes delayed detonation (DD) (Khokhlov 1991) 
     and pulsating delayed detonation (PDD), 
the initial subsonic burning front 
gradually makes a transition to a supersonic burning front (detonation). 
In the DD models, the detonation front erases the chemical structure that is 
left behind by the deflagration 
and creates a layered chemical structure, as observed. 
 With a particularly narrow range of energy variation,
 the DD models reproduce the optical and infrared light curves of typical SNe Ia quite accurately.
 In current 3-D pure deflagration models, however, 
 a significant fraction of the C/O WD remains unburned and 
 mixes with the burned material at all velocities,
in disagreement with observations.

In the second scenario, or DET2env models, 
combustion occurs in 
 a rotating low-density WD that is surrounded by an extended C-O envelope. 
The configuration forms during the merger of two low-mass binary WDs  
which lose angular momentum to gravitational radiation. 
The combined mass of the system may exceed the Chandrasekhar limit, but neither WD is above the limit. 
The third scenario is a double detonation of a sub-Chandrasekhar mass C-O WD, 
or the He detonation (HeD).
 The explosion starts with the detonation of a helium layer that has accreted from a close companion,
 which subsequently triggers the detonation of the inner C-O core.
 The exponential model is a good simplification 
 of the density structure obtained from various 1-D explosion models, as presented in Fig.1 of DC98.
          Both the W7 model and the delay detonation model DD200c 
           yield a perfectly exponential decline in density distribution,
     and they predict similar properties (HKW98).
            The PDD and the HeD (PDD3 and PDD1c; HeD10 and HeD6)  
            also exhibit more of an exponential than a power-law density distribution,
            although the gradient appears to drop in the innermost regions. 
         DC98 commented that the flattening 
    may have been exaggerated by the numerical effect of inadequate zoning near the center.
In the PDD and some of the HeD models, the explosion energy is below $10^{51} \erg$. 
  As low-energy models tend to be associated with a high density gradient, 
 the density evolution of lower-energy models is more sensitive to numerical resolution.
 The merger scenario is believed to be responsible for only a small fraction of the SNe Ia population
 because the predicted large amount of unburned C/O at the outer layers is inconsistent with 
  the IR spectra.
     Though not considered in DC98, this scenario may likewise bear an exponential density decrease,
   since a pure thermonuclear runaway reaction 
   (as opposed to a subsonic burning) 
   would incinerate almost all of the WD to form $\Ni$, 
 which process would be inconsistent with the observations of SNe Ia that only $\sim 0.6 \msun$ of $\Ni$ is generated (Hoflich 2003).

The chemical structure of SNe Ia varies with the explosion mechanism (HK96; DC98; HWK98; Hoflich 2003).
 In the accretion scenario,
 thermonuclear runaway takes place near the center,
 such that neutron-rich isotopes such as Fe, rather than $\Ni$, are synthesized in the core.
          However, for a model such as DD200c (HWK98), 
          the central void of $\Ni$ amounts to only
          $\sim 10\%$ of the total synthesized mass.  
The W7 model, whose density structure can be used as a prototype for an exponential model,
shows that the amount of Si substantially exceeds that of Fe in the denser parts of the shocked interaction region, 
while the Fe dominates close to the reverse shock, for the age and ambient density of Tycho's remnant (DC98). 
A similar composition is produced in delayed detonation or pulsating delayed detonation models for Tycho's remnant, 
except that the densest parts of the shocked ejecta appear to consist mainly of C and/or O. 
            In the merger scenario, by contrast,
            burning occurs in the low-density regions, causing little neutronization; 
             therefore, $\Ni$ is synthesized in the innermost region.
   In the HeD scenario,  two separate layers of $\Ni$ are generated,
       The outer one is formed by burning in the He envelope 
       above the velocity range 11,000-14,000 $\kms$.
          The mass of the inner $\Ni$ can be as little as that of the outer $\Ni$,
         or it can be considerable 
        and hence represent a bright SN Ia such as, for example, model HeD10 of HK96. 
In the models HeD6 and HeD10,
 the outermost dense regions of the shocked gas comprise mainly Fe,
 with O dominating in a dense shell inside Fe.
       The HeD scenario was advocated 
      in view of the large population of low-mass stellar progenitors.  
        However, Hoflich \& Khokhlov (1997) 
         maintained that 
           its spectra are generally too blue    
           to reproduce the observed light curves and late time spectra.

 Subluminous models giving acceptable fits to the observed SNe Ia; 
  for example, those in HK96 using a sample of 26 SNe Ia  
  tend to eject similar amounts of Fe along with the $\Ni$. 
 In such models, the initial mass of $\Ni$ ranges from 0.1 to 0.18 $\msun$,  
 while in normal models, it ranges from 0.49 to 0.83 $\msun$.   
   The properties of the Ni bubble expansion
  are dominated by the initial velocity distribution of $\Ni$ 
   and the effect of thermal pressure is negligible (\S~3).
   Therefore, including a realistic ejecta elemental distribution in our models 
      equivalently reduces the initial density contrast of $\Ni$,
  which is assumed to be distributed over the whole SN interior,
  with a mass fraction equal to $\omega$ everywhere within the bubble, 
  such that $\omega < 1$. 
  The range $0.5 \la \omega < 1$ is presumed.
  Such a simple parameterization of the $\Ni$ distribution 
   may sufficiently reproduce results based on a more detailed $\Ni$ distribution (\S~3.4). 

Figure~1 of DC98 shows that
the explosion models significantly diverge from the exponential models 
   as the velocity exceeds $\ga 10,000 \kms$,
   which transition presumably marks the burning interface between the core and the unburned envelope.
    Yet, inside the interface,
    the drop in density appears very smooth, 
   and the results suggest that
   the Ni bubble is not capable of expanding close to this region.

  To describe the evolution of an SNR in an exponential model, 
	   a set of scaling parameters, $R'$, $V'$ and $T'$, are used
	   following DC98 and W01:
	 \begin{equation}
	 R' = \left( {3M \over 4 \pi \rho_{am}}\right)^{1/3}  \approx
	 2.19 \ \left({M\over M_{ch}}\right)^{1/3} \  n_{am} ^{-1/3} \ \ {\rm pc},
	 \end{equation}
	 \begin{equation}
	 V' = \left({2E\over M}\right)^{1/2}  \approx 8.45 \times 10^3 \ \left({E_
	 {51} \over
	 {M/M_{ch}}}\right)^{1/2}
	 \ \  {\rm km \  s^{-1}},
	 \end{equation}
	 \begin{equation}
	 T' = {R' \over V'} \approx 248 \ E_{51}^{-1/2}
	 \ \left({M\over M_{ch}}\right )^{5/6}
	 \ n_{am}^{-1/3} \ \ {\rm yr},
	 \end{equation}
	 where $\rho_{am}$ is the ambient ISM density of the SNR 
and $n_{am} = \rho_{am}/(2.34 \times 10^{-24})$ gm cm$^{-3}$,
	 suitable for an ISM with an H/He ratio of 10/1 by number. 
	 The dimensional variables $r$, $v$ and $t$
	 can be expressed in terms of the non-dimensional quantities 
	 $r'=r/R'$, $t'=t/T'$ and $v'=v/V'$. 
       Since the deceleration of the SNR is caused by the ambient ISM,
    a higher surrounding density is responsible for a later evolutionary phase of the remnant (larger $T'$).
	 Once the non-dimensional solutions are obtained,
	 they can be re-scaled to the dimensional solutions 
	 for a different set of $M$, $E$ and $\rho_{am}$.
	 Thus, one evolutionary sequence in the non-dimensional variables 
	 represents virtually all possible dimensional solutions. 

 	  Figure~1 plots
	  the evolution of the characteristic velocities 
	  of an SNR 
	   that arises from the interaction of an exponential ejecta
	   with a constant ambient medium, corresponding to Fig.~1 of WC01.
	    As the ejecta density profile becomes flatter,
	    the reverse shock, initially moving
	    outward in the stellar frame, begins to turn inward at $t' \approx 2.5 $.
	    It reaches the stellar center at $t' \approx 8$.
  Unlike in the power-law model in which the evolution is self-similar (Chevalier {\it et al.} 1992), 
  the exponential model gives rise to different deceleration rates as the remnant evolves.
	  Nevertheless, the flow velocities immediately below the reverse shock front do not substantially deviate
          from those in the power-law model (Fig.~12 of W05)
 	  within the epoch $t' \approx 0.1-2'$, 
          when the clump-remnant interaction is likely to begin.

        \section {NICKEL BUBBLE SHELL STRUCTURE AND EVOLUTION} \label{sec:evo}

  	\subsection {Methods} \label{subsec:met}

	 The computational methods used 
        in this paper are similar to those adopted in the author's previous study W05.
	   HD and RHD simulations were performed 
 	  based on the two-dimensional finite-difference code ZEUS2D (Stone \& Norman 1992; Stone {\it et al.} 1992). 
	 Heating due to a central core of $\Ni$ that has  
         a density contrast $\omega=1$ relative to the ambient ejecta was considered. 
    	 The ejecta were initially freely expanding with an exponential
	 density profile that is characterized by an explosion mass $M$ and energy $E$.  
	   The radioactive energy was deposited continually to the gas thermal energy (HD case) 
 within the bubble 
 as the bubble-shell interface was tracked in a uniformly-expanding grid in spherical polar coordinates,  
	  starting at $t_{0}=100 \seco$ after the explosion.
    The inner boundary of the grid had a simple reflecting condition.
    At the outer boundary, a non-zero gradient outflow condition was applied
    to preserve the ambient exponential profile and eliminate spurious shocks that were caused by the grid expansion.
   Basko (1994) and W05 described the radioactive power input.
  Since the estimated mean life of $\Ni$ is larger than that suggested in Firestone (1996),  
    $\la 2\%$ less radioactive energy than indicated by Firestone was accumulated
  in the radiation-streaming epoch $\sim 10^6 \seco$.
    A $\Ni$ abundance of $0.5 \msun$ yields an effective energy input of $\sim 6\times10^{49} \ \rm erg$,
 or 6\% of the initial kinetic energy.
  Therefore, 
   the increase in the velocity from local free expansion is typically only around $\la 3\%$.

The initial ejecta velocity at the bubble edge was determined from the initial $\Ni$ density contrast and age:
	  $U_{0} = {R_{0} / t_{0}} \sim \omega^{-1/3}$.
	 The background thermal pressure was distributed
	 according to $p=\kappa \rho^\gamma$, where $\gamma = 4/3$ in the HD case,
             and $\kappa = 6.1 \times 10^{14}$ (cgs units)  
  	  (derived from the change in entropy during nuclear burning of C-O to $\Ni$, W05).
     The late-time radioactive power input prior to radiation streaming dominated the pressure in the bubble.
       In the RHD simulations,
       the radioactive power was deposited into the radiative energy 
       and $\gamma=5/3$ for matter.
    In supernovae, since the dynamic time required for the ejecta to become freely-expanding is much longer than that required for the radiation field to be thermalized with electrons
        and for collisions between electrons and ions to occur,
        a black-body radiation field is well established (Shigeyama \& Nomoto 1990).
         To ensure proper evolution of the flow toward the optically-thin regime,
         full radiative-transport RHD is invoked, under the assumptions of
         local thermal equilibrium (LTE) and gray opacity.
    The opacity of the ejecta is dominated by electron scattering; absorption is negligible (Shigeyama \& Nomoto 1990). 
          A Thomson scattering cross section of $0.2 \ \rm cm^2/gm$ was used
          to calculate the scattering opacity, 
          which is suitable for a gas of completely ionized heavy elements.
         The time for radiative energy to diffuse out of the bubble
     depends on the extent of electron scattering and the column density of the shell. 
   	 Notably, full-power radioactive heating can not be achieved in the radiative case 
        because radiative diffusion limits shell expansion.

	\subsection {Hydrodynamic Simulations} \label{subsec:hd}

 	Figure~2 and 3 show 
	the ejecta structure and the evolution of the dynamic properties of the Ni bubble shell
	in the adiabatic case, with parameters  
	$M=1.4 \Msun$, $E=10^{51} \erg$, $M_{Ni}=0.5\Msun$ and $\omega=1$,
	in what is referred to herein as the standard model. 
	 At the commencement of the simulation $t_{0} = 100 \seco$,
	the bubble has the radius $R_0=5.18 \times 10^{10} \rm \ cm$, 
	velocity $U_{0} = 5.18 \times 10^{8} \cms$ and
	 ambient density $0.92 \gcm$ at its edge.
	 The initial background thermal energy density is 
$2.77 \times 10^{16} \ergcc$ 
	 in the center.
	%
	 By this stage, 
	 a total radioactive energy of $4.85 \times 10^{45} \erg$ has been deposited since the explosion,
	yielding an averaged radioactive energy density of 
 $6.89 \times 10^{19} \ergcc$ in the bubble.
	%
	%

	 The inflation of the bubble  
	 induces a strong forward shock 
	 behind which the ambient gas is compressed into a dense shell (Fig.~2).
	A notable feature of the structure is  
	 the drop in density in the ejecta substrate.
	 The density contrast across the shock front is seven, 
	as for a radiation dominated ($\gamma=4/3$) strong shock.
	 The inner edge of the shell is a contact discontinuity  
	  where the gas has been shocked and cooled for the longest time and 
	 so has the highest density.  
	To describe the shell acceleration,
	the shock radius is approximated as a power-law function of time, $R_{sh} \propto t^a$,
	where $a$ is the expansion rate 
	at the shock front,
 which is equivalent to the velocity contrast between the shock and the ambient freely-expanding ejecta,
         $(dR_{sh}/dt)/(R_{sh}/t)$.
	The expansion rate rises to the maximum $a \la 1.05$ around
	$10^{7} \seco$ and then falls to 1.0. 
	That is, the shock front first accelerates 
 	and then freezes in the local free expansion of the flow (Figs.~3(a) and (b)). 
	The shell cannot be slower than the free expansion.

	The shell is very thin; 
	at $10^{7} \seco$ it reaches a maximum thickness ratio of
	$\beta \equiv h_{sh}/R_{sh} \la 0.004$,
 	 where $h_{sh}$ is the thickness of the shell (Fig.~3(c)). 
 	A linear relationship holds between the thickness ratio and the expansion rate, $\beta \approx (a-1)/10$, 
       because the expansion rate is low. 
 	The Mach number of the shock, 
 $      {\cal M}=\left({dR_{sh}/ dt} -{R_{sh}/ t}\right)
        \left(\rho_0 / \gamma p_0\right)^{1/2}
        = \ (a-1){R_{sh}/ t}  \left(\rho_0 / \gamma p_0\right)^{1/2} $,
	is $\cal{M}$$\sim 87$ at $10^6 \seco$,
	where $\rho_0$ and $p_{0}$ are the preshock density and pressure, respectively.
	%
	%
	  The expansion proceeds 
	  until pressure equilibrium is reached,
          or in the radiative case until the shell becomes transparent to the $\gamma$-rays.
	  The shell's compression ratio to the preshock ejecta is 
 	  $\chi \ga 10$ (Fig.~3(d)),   
	  and is limited by numerical resolution.
	 The compression in the bubble interior does not drop with
	time, as it does in the uniform-density model; rather it increases to $\sim 0.9$ (Fig.~3(d)). 

        At $\sim 10^6 \seco$,
	the shell has a thickness ratio $\beta \ga 0.002$, 
	a velocity $\vf \sim 6,000\kms$, 
	and a swept-up mass $\Ms \sim 0.09 \Msun$.  
	At this time, the surface density of the shell
	 begins to drop below the mean free path 
          for the $\sim 1 MeV$ $\gamma$-ray photons, 
          $\sim \ 33 \rm \ g/cm^2$,  
          determined by Monte Carlo simulations of X-ray transfer in SNe Ia
          (corresponding to an effective opacity of $0.03 \cmsg$; 
Sutherland \& Wheeler 1984, Shigeyama \& Nomoto 1990).  
	 The gas is then transparent to the radiation. 
	 Although $75\%$ of the total radioactive energy is yet to be released (Fig.1 of W05), 
	 further deposition of the radioactive energy does not affect the dynamics of the flow,
 	 and the flow evolves to a free expansion.

 	\subsection{Variation of Initial Parameters - Adiabatic Cases}

The effect of explosion parameters and other initial conditions on the solutions is now examined.
A wide range of explosion parameters, such as those used in the 1-D explosion models of HK96 and HWK98,
have been studied, 
    under both luminous and underluminous conditions (Tab. 1).
     The clump strength is conservatively evaluated by approximating
  the shell's frozen-in velocity by the flow velocity of the contact discontinuity 
      when the shell's surface density starts to drop below $\sim 33 \ \rm \cmsg$.

A comparison with the uniform-density model (model 'flat') indicates that in the standard exponential model (model 'std'), 
   the initial amount of $\Ni$ was distributed at lower velocities, such that
  radioactive pressure within the bubble is larger.
     The shell reaches a higher expansion rate,
     and the density contrast and thickness ratio are therefore greater.
  Nevertheless, the expansion velocity and swept-up mass of the shell are lower (Figs.~3(e) and (f)).
In the adiabatic uniform model,  
        the frozen-in velocity is 
    $\vf \sim 7,500 \kms$ and the swept-up mass is $M_{s} \sim 0.12 \Msun$,
   while in the exponential model, $\vf \sim 5,800\kms$ and $M_{s} \sim 0.09 \Msun$.
The difference between the velocities of the radiative counterparts of the above two models
     should not exceed the difference between their maximum velocities, $\sim 8,800 \kms$ and $\sim 7,500\kms$.

       For a lower initial $\Ni$ density contrast $\omega=0.5$,
  the frozen-in age is slightly advanced,
       with   
    a lower expansion rate, a lesser shell mass $\Ms \sim 0.03 \Msun$  
  and a higher frozen-in velocity $\vf \sim 9,300 \kms$. 
   The drop in the expansion rate and the swept-up mass 
      follow from the lower radioactive pressure within the bubble. 
    The frozen-in velocity is higher because the $\Ni$ initially extends to higher velocities.
In the exponential model,
 the radioactive pressure dominated the thermal pressure since the initial stage of heating; 
 the thermal pressure thus has a negligible impact.
   In the uniform model, 
   reducing $\omega$ results in a greater swept-up mass (Fig.~3(f)),
   because our assumed thermal pressure equilibrium across the bubble   
          causes the initial thermal energy in the bubble to be higher when $\omega$ is lower.
         Notably, the $\omega=0.5$ standard model
        yields a higher shell velocity  
         than its uniform-density counterpart,
         because of the initially higher velocities of $\Ni$.

   Comparisons of various models show that models with a higher velocity scale height $v_e$,
  or a flatter density distribution, tend to yield a higher shell velocity,
   a lower expansion rate and a lower swept-up mass. 
              For example,
 	      the HeD10 model (in which the density decline is the least among all of the models) 
  	      yields a shell velocity that extends to $\vf \la 13,000 \kms$ 
 	      and a swept-up mass of $\Ms \la 0.02 \msun$,
	      whereas the PDD1c model (which has the greatest density gradient) gives 
	        $\vf \sim 2,500\kms$ 
 	      (or $\sim 3,500\kms$ at maximum) 
 	      and $\Ms \sim 0.16 \msun$.
The shell velocity is correlated with the initial velocity distribution of $\Ni$, or the velocity scale height, 
because the shell expansion rate barely exceeds the free expansion of the local ejecta. 
The expansion rate is generally expected to be under $a \la 1.07$.
%
The momentum of freely expanding ejecta that were swept up into the shell,
 obtained by integrating the momentum density over the velocity space from the initial
 fo final shell location, 
 is found 
 similar to the final momentum imparted to the shell. 
 Thus, only a small fraction of the final shell momentum results from the bubble acceleration.
     Figure~2 shows that  
    the density profile of the bubble interior deviates from an exponential model
    mostly in the regions that are close to the contact discontinuity,
    also because the expansion is weak and can be contrasted with that in the case of core-collapse SNe (Fig.~2 of W05).
Figure~3(h) plots the time evolution of the approximate power index, $n \equiv v/v_{e}$,
 at the contact discontinuity.
 The frozen-in shell velocity does not exceed about three times the velocity scale height of the model -
 except for the luminous HeD10 and low-$\omega$ models.

  The swept-up mass of the shell is not very sensitive to the initial abundance of $\Ni$. 
 In a Chandrasekhar-mass, low-energy model using $E_{51}=0.7$ and $M_{Ni}=0.49$, for example, 
  the swept-up mass is comparable to that of a similar model with half the amount of $\Ni$.
It also is approximately the same across several models with intermediate explosion conditions,
with a typical $M_{s}$ of $\sim 0.1 \msun$.

   Solutions to the underluminous models 
    are not sensitive to variations
    of the initial parameters including the initial $\Ni$ density contrast, 
        because their steep density distribution and low $\Ni$ abundance allow little corresponding variations 
         in velocity and radioactive pressure.
   A unique feature of the exponential models except for PDD1c is that 
   the density contrast at the bubble interior increases over time,
   because of continual flattening of the exponential density distribution and a relatively low expansion rate.

 In the HeD scenario, the outer layer of $\Ni$ is present at high velocities of
  above 11,000-14,000 $\kms$ (HK96).
 An off-center heating scenario was considered using the HeD6 model (model 'HeD6*'), 
   in which the radioactive power from a half of the outer $\Ni$ mass was gradually deposited in the space
   between the velocities $12,000 \kms$  
            and $\la 13,400 \kms$,   
   while the inner grid boundary continues to move at a fixed velocity of $12,000 \kms$.
             A dense shell as in the central-heating model is swept up, 
 but it exhibits very weak acceleration relative to the local fast ejecta.
      The onset of the optically-thin stage thus advances to $\sim 10^5 \seco$,   
      leading to a 'low' frozen-in velocity at $\la 14,000 \kms$.

 A flatter distribution of the thermal pressure, $p \sim \rho$, was assumed   
 	  with a constant temperature, $kT \sim 0.6 \ MeV$ (derived from the nuclear burning process, W05). 
          As the central background pressure is enhanced by a factor of $\sim 500$,
	 the radioactive pressure still dominates the thermal pressure,
 	   so the evolution of the Ni bubble remains largely unchanged.
      To account for the central $\Ni$ void as in the accretion scenario, 
       $\Ni$ in model DD200c was redistributed to higher velocities, 
         corresponding to a void mass of 1/10 of the synthesized mass.
       The result is indistinguishable from that of the central-heating model.
	Lastly,  
	the simulated solutions that began at the age $t_0=1000 \seco$ 
       rapidly converged with those that began at $100 \seco$.
Thus, as long as $t_0 \la 1000 \seco$,
the final solutions do not depend on the commencing ages of the simulations.

 \subsection {$\Ni$ Mass Fraction in the W7 Model}

 In our $\omega = 1$ exponential W7 model, the shell velocities, 7558 $\kms$,
is actually lower
 than the minimum velocity $\sim 10,000 \kms$ of the Si-dominant layer, obtained using the W7 model of Nomoto {\it et al.} (1984).
In their original calculations, the velocity of the radioactive Ni was as high as 8300 km/s,
and its fractional mass fell to 0.5 at 8800 km/s. 
   To justify that a 
   simple parametrization of the initial $\Ni$ distribution 
   in terms of $0.5 \la \omega < 1$  
   approximates the results that consider a realistic ejecta elemental distribution,
   the $\Ni$ mass fraction given by the model of Nomoto {\it et al.} ($M_{Ni}=0.58 \msun$) 
   was incorporated into the exponential model.
  As the $\Ni$ mass fraction falls steeply above $8,800 \kms$,
   the initial $\Ni$ distribution was cut off at this velocity but the same radioactive power,
    proportional to the mass density within it, was deposited.
 Spatial variations in the background thermal pressure 
 caused by the nonuniform elemental distribution were neglected.  
 This modified model
 gives a similar initial outbound speed to that obtained in our $\omega=0.7$ W7 model (Fig.~4). 
(The integrated density over
  all elements should follow an exponential profile as illustrated in DC98.)
In both models, the final shell velocity reaches $\sim 9000 \kms$.
In the modified model the swept-up mass is lower but does not differ significantly from that in the $\omega=0.7$ case.


The reason why the use of $\omega=0.7$ adequately reproduces  
the result based on a more detailed $\Ni$ distribution in the W7 model
is that it places the outer boundary of $\Ni$ 
in the 8300-8800 velocity range, beyond which whose mass fraction drops down rapidly. 
However, in view of a finite sound speed propagation within the bubble and the presence of nonradioactive
material in the ejecta, 
this result strongly depends on the assumed value of $\omega$ and is merely an approximation.
         On the other hand, the clump speeds in our $\omega=1$ cases 
         stand only as the lower limits in more realistic Ni bubble heating scenarios.

    \subsection {Radiation-Hydrodynamic Solutions} \label{subsec:rhd}

Here we present full-radiative transport RHD simulations 
to illustrate the effect of the diffusion of radiative energy across the bubble structure
 on the shell properties. 
     In Figs.~2, 4 and 5, 
 the density structures of the radiative models are plotted over their adiabatic counterparts.

      The shock has a widely extending radiative precursor,
        lowering the overall compression of the shell.
   When the radiative precursor is considered, the shell mass reaches $\sim 90\%$ of the adiabatic value (0.1 $\msun$).
   Excluding the radiative precursor reduces the mass to $\ga 0.04 \msun$ (radiative case) in the standard model. 
   The drop is typically $\la 50\%$.

 Since the acceleration of the shell is weak, 
   the shell formation processes can be regarded as perturbations to the initial free expansion, 
  such that the final radius and velocity of the shell are only slightly reduced.
 The final shell velocity differs from the adiabatic solution by  
  $\la3\%$ (Fig.~6). 
  As the final shell velocity may not be an adequate measure of the radiative effects, 
  the difference between the final shell velocity and the velocity of the initial free expansion  
   at the bubble's edge is used
  to assess the importance of radiation diffusion effects.
   A $\sim 25\%$ difference is found between the adiabatic and radiative solutions.
   However, in the $\omega=0.7$ W7 model, it is only 12\%. 
   Hence, the radiative effects are expected to be mild in a more
    realistic scenario.

     The mass and number of clumps should drop   
     as the acceleration and compression of the shell are decreased.
     Since the RHD simulations are affected by inadequate spatial resolution, 
     the shell's density contrast is crudely estimated, 
     based on the assumption that
     that the shell mass is 50\% less and the shell thickness is 
 double the size of  
    that in the adiabatic solutions;
       a compression ratio of $\sim 25\%$ of the adiabatic value, or $\chi \sim 25$, is expected.
     Nonetheless, since ejecta clumps (shell fragments) most likely arise from the densest 
     part of the CD,   
     the true compression ratio may be higher.


	\section{SELF-SIMILAR APPROXIMATIONS}

    To provide insights into the exponential model,  
      this section discusses the self-similar solutions 
	of Chevalier (1984) and Jun (1998) 
	for a pulsar bubble that arises from a pulsar wind that interacts with
	 supernova ejecta. 
  The analogy to the problem of interest follows from the fact that
    the shock is driven by the motion of CD, as in the process by which a piston
  shoves up mass in a gas chamber. 
  Therefore, the properties of the forward shocked layer (shell) can be determined by the motion of the piston (CD) and 
  the density of the gas downstream (outer ambient ejecta);  
  details of how the inner gas is heated to power the piston need not be known to apply self-similar solutions.

   The Chevalier (1984) and Jun (1998) solutions 
        considered a pulsar wind with a time-varying luminosity $L \propto t^{-l}$
        and supernova ejecta with 
 a constant power index $n$ associated with the denstiy decline.
 	A simple dimensional analysis reveals the self-similar expansion rate of the bubble to be 
          $a_{ss}={(6-n-l)/(5-n)}$.
	 Accordingly, the rate of expansion increases with the density gradient.  
	The self-similarity exists only for
	 $a_{ss} \ga 1$, or $l \la 1$,
	since the shock cannot be slowed down in the ejecta to be slower than the free expansion,
	and for $n \la 3$,
	since the mass must be finite.
	With the given adiabatic index $\gamma=4/3$,
	self-similar solutions can be described by two parameters: $a_{ss}$ and $n$. 
Since the exponential density gradient increases with radius and flattens over time,
         $n$ is evaluated
         at the contact discontinuity near the frozen-in age of $10^6 \seco$ (Fig.~3(h)).
        In our standard case, $n=2.4$.

   In the evolution, the shock is expected to accelerate ($a > 1$) 
    to the self-similar regime, 
   and so the shell thickness rises continuously.
   If a constant luminosity $l=0$ and $n=2.4$ powers the expansion, 
then $a_{ss}=1.38$.
	   In our hydrodynamic solution (simulated solution of the standard model),
	 the shell thickness ratio reaches a maximum at $5 \times 10^6 \seco$; this is correlated with 
	 an increase in the accumulated radioactive energy 
	  at $5 \times 10^6 \seco$ (Fig.~4 and 9 in W05).
	The turnover after $10^{7} \seco$ follows from the fact that $l>1$  
	 (where $l$ is estimated by taking the time-derivative of the accumulated radioactive energy,
	  $l \sim -dlnq(t)/dlnt$, where $q(t)$ is as given in W05),
	 when the self-similarity breaks down;
	 it is not related to the mass of $\Ni$ or the gradient of the ejecta.
	  The decrease in the shell thickness during the earliest evolution 
	  is due to the negligibility of the medium motion during a sudden acceleration
            (Jun 1998),
	   which can be regarded as a numerical artifact that depends on the starting age of the simulation.

      The hydrodynamic solution has a weaker expansion than the self-similar solution for a fixed density gradient $n=2.4$,
           but gradually approaches the self-similar regime.
   In the self-similar solution, $a_{ss}=1.07$ at $1 \times 10^{6} \sec$ 
   (for $l \sim 0.8$, which is the luminosity index at $1 \times 10^{6} \sec$), 
  and $a_{ss}=1.0406$ at $2 \times 10^6 \seco$ (for $l=0.896$), 
    while in the hydrodynamic solution, $a=1.04$ and $1.0444$ at the respective ages.
     The self-similar solution thus approximates to the hydrodynamic solution near the frozen-in stage 
    (Fig.~7). 
  The self-similar density is infinite at the contact discontinuity.
Chevalier (1984) gave 
 the asymptotic density profile close to the contact discontinuity in the uniform model: 
  $\rho \propto (r/R_{CD} -1)^{-b}$, where $b=2a_{ss}/(7a_{ss}-5)$.
This result can readily be applied to a power-law model.
Because of the sharp increase in density toward the CD,
  the coarseness of the grid obviously limits the highest computed density in the grid domain.

The self-similar solutions to the thickness ratio $\beta$ and average density contrast $\chi_{avg}$ of the shell
  are derived as a function of the self-similar expansion rate $a_{ss}$ (Fig.~8).   
         Both $\beta$ and $\chi_{avg}$ increase with the density gradient.
  $\chi_{avg}$ reaches a maximum (turnover) with respect to $a_{ss}$ in cases of $n < 3$.
 For example,
 for $n=2.4$, 
  the density contrast turns over at $a_{ss}=1.04$ and the shell thickness ratio
  is about 0.004.
  $\beta$ is under $\la 0.023$
   when the largest expansion rate allowed for an expanding medium, $a_{ss}=(6-n)/(5-n)$, is achieved.  
 (The Chevalier and Jun solutions used gamma=5/3 and thus yielded a larger thickness ratio, about 0.02 for $n=0$ and $a=1.2$.)
 In the $n=2.4$ case,
 $\beta$ ranges from $\sim 0.0055$ to $\sim 0.006$ for $a_{ss} \la 1.06$,
        and $\chi_{avg} \sim 75$,  
          depending only mildly on the expansion rate between $1.02$ and $1.08$.
         $\chi_{avg}$ may be enhanced by a factor of $\sim 50\%$ over that in the $n=0$ case. 
      Figure~7 displays a larger shell thickness in
  the hydrodynamic solution than in the self-similar solution.
       If the hydrodynamic solution reaches an equally infinite density at the CD,
    then the true average density contrast of the shell may be even higher than that given by the self-similar solution.

 	We caution that, 
        the self-similar solution can also be affected by the spatial resolution 
 	used in integrating the governing self-similar differential equations 
       from the shock front to the CD. 
        High-resolved results are generated using the fourth and fifth -order Runge-Kutta methods 
         with a small integration step size
        until the density contrast values do not increase further (Fig.~8).
  Fluctuations in the curves are 
  caused by the accumulation of errors in summing up the shell mass over
  the acquired density distribution.

       We have shown that the hydrodynamic solution of an exponential model
     can be approximated by the self-similar solution of a power-low model. 
        Nonetheless, such a description is merely 
          semi-quantitative for the problem under consideration,
        as it has no true self-similar solution.

	\section{NICKEL BUBBLE SHELL AS EJECTA CLUMPS} 

       This section examines whether the ejecta clumps that originate in the Ni bubble shell
       can initiate the clump-remnant interaction sufficiently early in time
       and further survive crushing in a remnant,
        such as Tycho's knots, under various explosion conditions (Tab.~1).  
        As a shock moves past a cloud,
 the incident shock generates a transmitted 'cloud shock' that moves into the cloud and crushes it.
       When the cloud shock exits the cloud, 
       a rarefaction wave moves back into the cloud and causes lateral expansion.
  The combined instabilities 
  destroy the cloud on a timescale of several times the cloud-crushing time,
 $t_{cc}=r_{c}\chi^{1/2}/v_{0}$, 
 where $r_{c}$ is the cloud size and $v_{0}$ is the initial shock velocity (Klein {\it et al.} 1994).
 	When a clump interacts with a remnant, 
  the clump survival depends on three parameters:
  (1) the initial density contrast of the clump, $\chi$, relative to the surrounding ejecta;
  (2) the initial impact age of the clump with the reverse shock, $t_{RS}$;
	 this is related to the absolute ejection speed of the clump,
	 which is given by the frozen-in velocity of the Ni bubble shell,
	as well as the true age of the remnant, 
	and (3) the initial size of the clump. 
    Of the three parameters, the clump speed and initial impact age with the remnant
    are barely limited by resolution
and these properties are well-defined for comparing observations.
     As they do not significantly diverge in the case of radiative diffusion,
    the following discussions are primarily based on the adiabatic solutions.

  \subsection {Tycho's SNR as the Standard Model}

Tycho's remnant is used as a prototype for the clump-remnant interaction.
 Table~1 lists the speed of the clump (frozen-in velocity of the shell) 
  for each model under consideration.
The initiating age of a clump-RS impact 
  when the clump speed matches the flow velocity immediately below the RS is derived 
using the velocity evolution of a remnant in Fig.~1,
The dimensional values of the ages are noted specifically 
 for two ISM density values, $n_{am}=0.5$ and 1. 
  The final column of Table~1 presents compares of the normalized age of Tycho's remnant (whose real age is 435 yrs) with the RS-impact; 
   the comparison yields favorable explosion conditions for Tycho's supernova.

    Since the remnant of a Type Ia SN continuously decelerates with time,
 its explosion parameter and ambient density can be inferred from 
 the observed deceleration rate of the remnant, which parameter is independent of distance, given its dimensional age.
For a Chandrasekhar-mass, $E_{51}=1$ explosion,
DC98 noted that the observed deceleration rate $\delta \sim 0.5$ 
(DC98 and references therein; Hughes 2000) of Tycho's remnant
depends on an ISM density in the range of $n_{am}=0.6-1.1 \cmc$. 
     In this case, Tycho's dimensionless age is between
     $t'\sim 1.39$ (for $n_{am}=0.5\cmc$) and $\sim 1.75$ (for $n_{am}=1.0\cmc$).
       For $n_{am}=1$,
      the reverse and forward shocks have radii of 2.08 pc ($r' = 0.98$) and 3.09 pc ($r' = 1.45$),
and decelerate with
       $\delta = 0.15$ and $\delta = 0.47$, respectively;
            the result is consistent with the observations.
   However, since the deceleration of the forward shock evolves very slowly 
  near Tycho's present age, 
 $t' \sim 2.0$, when $\delta \sim 0.5$ (inset of Fig.~1),  
   the explosion parameters can be rather unconstrained.

   \subsection{Clump Speed and RS-Impact Age - Indication of Explosion Conditions}

   As our standard $\omega=1$ model yields a lower clump speed than the uniform model (for which $t'_{RS} \sim 0.9$),
  the onset of a clump-remnnat interaction is delayed  
to the age $t'_{RS} \sim 1.3$, or $\sim 330$ yrs, for $n_{am}=1$ 
 (see Fig.~1).
    The clump-RS impact still precedes Tycho's present epoch,
   but the clump-FS impact does not, and is anticipated to occur at 
   $t'_{FS} \sim 2.2$, or $\sim 545$ yrs 
   (Fig.~9).
   The clump speed is $\sim 30\%$ short of the velocity of Tycho's knots,
$v \sim 8,300 \kms$,
    estimated at a distance of 2.5 kpc. 
    Nonetheless, measurements on Tycho's distances range from 1.5 to 4.5 kpc (see Schaefer 1996, Ruiz-Lapuente 2004 and references therein),
  so the observed velocity of the knots can also be quite uncertain.

Explosion conditions similar to 
  those in the DD200c and W7 models 
  result in a timely clump-remnant interaction as well as a 
 compatible clump speed. 
The luminous HeD10 model gives a high clump speed but is not favored 
  as its clump-remnant interaction
 ($t'_{RS}=0.56$ and $t'_{FS} \sim 1.0$)
  may be initiated too early to be sustained to Tycho's current state. 
  The HeD10 model is expected to yield a weaker clump strength than the typical models 
  because of its lower shell expansion rate and density contrast.

 The underluminous PDD1c model yields a low expansion velocity $\sim 2500 \kms$ 
  where the exponential model becomes invalid (Fig.~1 of DC98).
If a constant ejecta density of $10^{-12} \gcm$ at $10^{6} \seco$  
or $1 \gcm$ at $t_{0}=100\seco$ is assumed in this velocity space,
then the initial bubble velocity is $3600-4500 \kms$ for $\omega=1-0.5$,
and the final shell velocity may be $4000-5000 \kms$, which 
 is still insufficient for Tycho's knots.
 For models with a moderate clump speed, the high-mass DET2env6 model
 notably gives a relatively late clump-RS impact, 
 $t_{RS} \ga 400$ yrs, for $n_{am}=0.5\cmc$, 
 while the low-mass HeD6 model yields a timely interaction at $t_{RS} \sim 270-340 \rm \ yrs$ 
 with a similar clump speed,
despite a low $\Ni$ abundance $M_{Ni}=0.25 \msun$. 
The uncertainties in the observed clump speed and deceleration rate of the remnant 
suggest that Tycho's SN may possibly have originated in the HeD6-like, sub-Chandrasekhar mass and low-energy explosion scenario.

 As the density of the SN ambient medium declines (e.g., $n_{am} = 0.5 \cmc$, see Tab.~1), 
   the clump-RS impact is further delayed, 
   and the clump undergoes a longer intershock-crossing process 
 (Fig.~10). 
 The lower density limit $n_{am} = 0.6 \cmc$ 
 for Tycho's deceleration rate 
 in DC98 could however be underestimated,
   as the undecelerated motion of clumps behind the blast wave was not considered.
 Given that the dimensionless time scale $T'$ increases by a factor of 2.15
  for a ten-fold drop in ambient density, 
   if the minimal ambient density is $0.05 \cmc$,
then the RS impact should occur within an age of $\sim 2000$ yrs under all plausible explosion conditions of SNe Ia.

    In the case of an initially lower $\Ni$ density contrast $\omega$, 
   a faster but smaller and less dense clump is created. 
Consider the standard $\omega=0.5$ model as a reference:
 the cloud-crushing to intershock-crossing time ratio reaches only $\la 2 \%$;
         the shell to Si knot mass ratio falls to $\sim 1/13$,
         and the forward shock impact advances to $t'_{FS} \sim 1.0$.
 The clump speed (approximately the speed of Si) obtained using
Nomoto's W7 model may be approximated by our $\omega=0.7$ W7 exponential model.
  The $\omega=0.5$ cases are particularly applicable to underluminous explosions 
 which tend to eject an amount of Fe  
that is comparable to that of Ni.
    However, since the clump speed in such scenarios depends less sensitively on variations of the initial conditions,
    the increase in the clump speed relative to the $\omega=1$ case does not suffice to change our conclusions.

 \subsection{Clump Density and Size - Indication of Clump Robustness}

 The clump is expected to be of a size that is comparable to the thickness of the Ni bubble shell.
         In Fig.~9,  
         the fractional clump size is estimated versus the age
 at which the clump-remnant interaction is initiated, for varying thickness ratios ($\beta$).
   	For $\beta=0.004$, 
	the clump initially occupies 
          $\sim 1\%$ of the intershock shock width, 
         or $\sim 0.3 \%$ of the forward shock radius, 
    upon the time of the RS impact.   
     If the lateral expansion of the clump is neglected,
     the uniform expansion of the clump 
   (in which the density falls as $\sim t^{-3}$ and the size increases as $\sim t$)    
    would result in a fractional clump size in terms of the forward shock radius
   that equals the initial thickness ratio of the shell, 
    when the clump reaches the forward shock.
%

After they enter the SNR intershock region,
the clumps become fragmented over a period of several times the cloud-crushing time
   (WC02, equation (8)). 
    The self-similar solution 
    indicates that the shell's average density contrast is up to $\sim 50\%$
    greater than that obtained using the uniform model.  
 The Rayleigh-Taylor mixing at the Ni bubble shell interface is thus facilitated, 
 creating denser clumps.
The cloud-crushing time scale of the clump in our optimal adiabatic scenario 
$\chi \sim 100$ and $\beta=0.004$ is estimated to be 
$t_{cc} \ga 20$ yrs, 
or $\sim 0.06$ times the intershock-crossing time $(t_{FS}-t_{RS})$. 
Although the time scale seems small, 
 in the high-resolved simulated clump-remnant interaction (WC02),  
  clumps fragments successively create protuberances on the forward shock in the remnant, are decelerated, and the forward shock front catches up with them; 
 the clump crushing process is not expedited. 
  The dynamic effect of small but dense clumps thus is not negligible.
Notably, for a fixed clump mass, the cloud-crushing time scales inversely with the clump size, 
$t_{cc} \propto r_{c}^{-1/2}$.

  Recent Chandra observations of Tycho's SNR by Warren {\it et al.} (2005) revealed 
   the presence of dense gas that stretches very close to the blast wave - 
much closer than the unstable gas according to the exponential model of WC01. 
           Warren {\it et al.} attributed such a structure to the intershock unstable gas 
          that arises from a different intershock density profile,
        modified by the cosmic ray accelerations of the preshock gas.  
 However, 
  the small and dense clump fragments can also contribute to the observed dense gas;
   their interaction with the remnant does not extend the CD to larger radii. 
  Additionally, modification of the intershock structure
  should have a weak dynamic influence on the dense clump fragments.

\subsection{Clump Mass - Indication of Ni bubble Instability}

        {\it ROSAT} observations of Tycho's remnant have revealed 
          two X-ray knots that protrude from the SE edge of the remnant in undecelerated motion.
    Hwang, Hughes, \& Petre (1998), 
         estimated a mass of $0.002 \msun$ and $0.0004 \msun$ for the Si+S and Fe knots,
         respectively. 
 The total swept-up mass in the Ni bubble shell ($M_{s} \sim 0.1/0.05 \Msun$ in the adiabatic/conservative radiative case) 
is then
 $\sim 25-50$ times the mass of the Si knot.
 While the process by which the shell fragments are formed is yet to be demonstrated, 
 this mass ratio could be suggestive of the instability mode at the bubble-shell interface, 
 where the density and pressure exhibit opposite gradients and thus satisfy the Rayleigh-Taylor unstable criterion 
 (Chevalier 1976).
  The instability growth time can be inferred from the characteristic relaxation time $\tau$ 
of a perturbation, given the differences across the contact discontinuity, 
in the acceleration (effective gravity) and 
position, $\Delta g$ and $\Delta x$: 
  \begin{equation}
 {1 \over \tau^2} \equiv {\Delta g \over \Delta x} 
 =  {1 \over \Delta x}{\partial P \over \partial x}(\Delta {1 \over \rho}).
  \end{equation}
By the epoch of radiation streaming $10^6 \seco$,  
  the shell has been accelerated over three orders of magnitudes in radius 
and $\tau \sim 10^4 \seco$, so
   the instabilities are estimated to have fully developed.
 Alternatively,
  the growth of the instability can be evaluated at any stage 
   using the information on the shell acceleration and thickness contained in Figs.~2(b) and (c),
  given that
  the growth rate of an instability mode that initially disrupts the shell 
   is related to the square root of
 the ratio of the shell acceleration to the shell thickness, or
$ (t/\tau)^2 = (da/dlnt + a(a-1))/\beta. $

%

    To determine the the spatial structure of SNe Ia, 
      Dietrich Baade measured 
    the polarizations in the Si II 635.5 nm line,
  which is a very common feature among young SNe Ia up to an age of $10^6 \seco$  
(2006, private communication). 
He concluded that
in order to generate the observed polarized effect of SNe Ia,
there must exist a few tens ($\ga 10$) of clumps or bubbles in the ejecta, 
and their number must not be too large.
  However, 
 he also noted that 
 the asymmetry produced in multi-dimensional SNe Ia explosion models 
yields approximately the same view in all directions  
 (Reinecke {\it et al.} 2002; Schmidt \& Niemeyer 2005),
   so that none of the models can account for the polarization.
  To the contrary, Baade's estimate of the number of spatial inhomogeneities of SNe Ia is consistent with 
  the shell-to-clump mass ratio herein. 
    This result indicates that the polarization may originate in the Ni bubble effect,
    which is characterized by a weak expansion.

\subsection {Indication for Tycho's SN, SN 1006 and N103B}

Indications that both Tycho's SN and SN 1006 were underluminous
and corresponding indications for SNR N103B are finally examined.

Van den Bergh (1993) reconstructed the light curve of Tycho's SN, which 
  he found was very similar to that of the underluminous SN 1991bg (Filippenko {\it et al.} 1992; Leibundgut {\it et al.} 1993).
 Photometric calibrations and estimates of distance to Tycho's SN vary widely, 
and recent interpretations of the historical records 
by Ruiz-Lapuente (2004) identify Tycho as a normal Type Ia supernova. 
For SN 1991bg,
Turatto {\it et al.} (1996) estimated an explosion energy that was a factor of three to five below that of typical SNe, 
 and Mazzali {\it et al.} (1997) suggested 
   a WD mass of $0.62 \msun$, a kinetic energy of $E_{51}=0.62$ and an $\Ni$ mass of $0.07 \msun$
  to account for the extension of the observed centrally-peaked $\Ni$ to $5,000 \kms$. 
  Recent SN surveys (Li {\it et al.} 2001; Benetti 2005) classified  
     as many as 36\% of the observed SNe Ia as being intrinsically peculiar spectral types: 
20\% were like the overluminous SN 1991T (Mazzali {\it et al.} 1995),
while the remaining 16\% were similar to the underluminous SN 1991bg. 
As a conservative approach to understanding the inferred underluminous condition,
 a Chandrasekhar-mass and low-energy model with $E_{51}=0.7$, $M_{Ni}=0.25$ and $\omega=1-0.5$ were examined. 
The assumed Ni abundance is roughly the upper mass limit of Fe ($\la 0.15 \Msun$) in SN 1006 (Hamilton {\it et al.}
    1997).
     In this PDD-like scenario, the shell expands at $v_f \sim 3,800-5,000 \kms$;  
      the velocity is not consistent with Tycho's SNR. 
    The parameters of Mazzali {\it et al.} (1997) were then adopted;
    in this case, $v_f \sim 4000-5000 \kms$ for $\omega=1-0.5$ - 
      roughly comparable to the observed velocity of $\Ni$ in SN 1991bg.

The peak magnitude of SN 1006 has been variously estimated at values that span over five magnitudes.
More recent examinations by Winkler {\it et al.} (2003) 
imply a normal brightness,
 but the Fe abundance of the remnant is known to be peculiarly low (Hamilton {\it et al.} 1997), 
revealing the SN likewise to be underluminous.
The morphology of the SN 1006 remnant is not symmetrical;  
its ambient density may be three times higher toward its northwest region 
  while the overall observed deceleration of the remnant is $\delta \sim 0.5$.  
DC98 proposed 
 a Chandrasekhar-mass model of $n_{am} = 0.047 \cmc$ for $E_{51}=1$,
or $n_{am} = 0.07 \cmc$ for $E_{51}=1.3$ (model 'W7-2') to explain the deceleration. 
 With $\omega=0.7$ and $M_{Ni}=0.59$,
both models predict a reverse shock impact before an age of 500 yrs and a forward shock impact
in the present epoch.
With $\omega=0.5$ and $M_{Ni}=0.16$,
   an expansion velocity at $\vf \sim 5000 \kms$ is yielded; 
   this value is consistent with the lower observed velocity limit of the Si in the remnant,
  $5,600-7,000\kms$, estimated for a distance of 1.8 kpc. 
   The distance measurements of SN 1006 range from 0.7 pc 
   to $\sim 2.5\ \rm kpc$ (see DC98, Winkler {\it et al.} 2003 and references therein).
In the case of $v_f \sim 5000 \kms$ and $n_{am}\sim 0.05 \cmc$,
the clump-RS impact is expected at an age of $\sim 1000$ yrs.
   This result indicates that the clump-remnant interaction may have just started in SN 1006.

The exponential model explains 
the commonly-observed lack of mixing in the remnants of SNe Ia.
Blondin {\it et al.} (2001a) studied the interaction of numerous small ejecta Fe bubbles 
with the intershock region of an SNR.
As the reverse shock front moves back more rapidly through the bubbles,
 considerable turbulence is generated,
   which eventually destroys 
the original stratification of the ejecta.
In their simulations, a density contrast of $\chi \sim 1/100$ was used between the bubbles and the ambient ejecta.
 Although the assumption of high density contrast pertained specifically to core-collapse SNe 
 because of the expected initial mixing of $\Ni$,
 such a turbulent scenario is unlikely to apply for SNe Ia, 
    because the bubble interior has a density contrast that is close to unity, 
 and clumps are expected to be present only in a thin shell at the boundary of the Ni bubble. 

A good example of the Ni bubble effect is the young and compact supernova remnant N103B
located in the Large Magellanic Cloud
(Lewis {\it et al.} 2003).
 In this remnant, the outer Si and S emissions exhibit an ionization time scale ($n_e t$)
    that is $\sim 100$ times higher than that of the interior, hot Fe,
  while the overall chemical composition is surprisingly stratified   
  without significant mixing among Fe, Si and 
  other ambient ejecta gas.
Light echo (Rest {\it et al.} 2005) and X-ray observations (Hughes {\it et al.} 1995)
yielded estimates of the age of N103B of $\sim 850$ and $\sim 1000$ - 2000 yrs, respectively.
Given the age,
the clump-remnant interaction should have started in N103B.
The consistency between the ionization time scale ratio of Si to Fe 
and the density contrast across the Ni bubble shell demonstrated that the Ni bubble expansion had occurred,
 and yet the globally stratified composition 
 reveals suppression of mixing in the clump-remnant interaction.
 This outcome is further consistent with our expectations of the exponential model.

\section{SUMMARY AND CONCLUSIONS}\label{sec:conclu}

We have investigated the structure and expansion properties of the Ni bubble shell in SNe Ia
due to radioactive heating
caused by $\Ni \rightarrow\ \co \rightarrow\ \fe$ decay sequence,   
assuming an exponentially-declining density profile of the ejecta substrate.
 Based on the indication that ejecta clumps arise from 
 the breakup of components of the Ni bubble shell,
 whether the inferred shell properties are compatible with those
 revealed for the ejecta knots in Tycho's remnant was studied.  
  Adiabatic solutions were first obtained, and then radiative solutions, 
 including the radiation diffusion process across the Ni bubble shell, 
  up to the approximate frozen-in age $\sim 10^6 \seco$,
   were applied to estimate the clump properties. 
 Since a realistic elemental distribution of $\Ni$ results in a higher expansion velocity than
 a centralized distribution of $\Ni$ used in our model,
 which effect is equivalent to the use of a lower initial $\Ni$ density contrast in 
 the range of $0.5 \la \omega < 1$, 
  our $\omega=1$ cases may only hint at the lower limit on the clump speeds.
 We believe that our inferred clump strength is at least not overestimated.

 The expansion of the Ni bubble sweeps up a dense thin shell of
 $\sim 0.1 \Msun$ shocked gas in a typical Type Ia SNR.
 The density of the shell increases inwardly toward the bubble-shell interface, 
 such that the highest computed density is limited by numerical resolution.
 Since the exponential profile is close to
   power-law profile that evolves 
   from a steep power law to a flatter one,
the adiabatic solutions of the shell were approximated as self-similar solutions for a pulsar bubble shell 
 in power-law ejecta of power index $n$ 
   with the shock propagated at an expansion rate $a$, 
  which in our intermediate-class models falls in the range $n=2-3$ and $a\la 1.07$. 
   The shell is thicker than given by the flat ejecta model, 
   representing $\la 1\%$ of the forward shock radius 
    and the mean density contrast is enhanced by $50\%$.
In the radiative case, 
  the shell is broader and less dense; 
the total swept-up mass of the shell is $\sim 50\%$ less without the radiative precursor,
  but the shell's expansion velocity is only $< 3\%$ lower close to the frozen-in age.  
The radiative diffusion process does not substantially affect
the clump speed or the initiative age of the clump-remnant interaction.

  A wide range of explosion parameters that are
  similar to those used in valid 1-D explosion models was explored. 
    The exponential model gives a higher shell expansion rate and thickness ratio than the uniform model, 
    and a lower shell velocity and swept-up mass. 
The expansion of the shell is only slightly
 faster than the free expansion of the ambient ejecta substrate.
 Consequently, the frozen-in velocity of the shell
  often rises with the velocity scale height $v_e$ of the exponential profile -- which characterizes the free motion of the ejecta substrate -- 
 or drops with the ambient ejecta density gradient.
  The velocity and swept-up mass of the shell appear to be inversely correlated: 
 steeper models (with smaller velocity scale heights) tend to acquire
   a lower expansion velocity and a higher swept-up mass.

 Under intermediate explosion conditions  
  with an ambient ISM density in the range of $n_{am}=0.5-1\cmc$,  
 the clump-remnant impact is anticipated to occur at $\sim 300$ yrs after the supernova is formed. 
Although the impact precedes Tycho's present state,
the clump speed ($\sim 6000 \kms$ for $\omega=1$)   
  may not be sufficient. 
 A higher clump speed and an earlier clump-remnant impact is expected   
 when the model incorporates a more realistic elemental distribution.
The explosion parameters obtained from the successful explosion models 
such as the deflagration W7 and the delay detonation DD200 
yield the most favorable result for Tycho's knots.
 A sub-Chandrasekhar-mass, low-energy scenario such as the He detonation HeD6 
 may also satisfy Tycho's condition.  
 A Chandrasekhar-mass and low-energy model such as the PDD 
 is unfeasible,
 indicating that if Tycho's SN was intrinsically underluminous, 
 then a near-Chandrasekhar-mass progenitor could not arise.
Likewise, SN 1006 could not originate in a W7-like scenario 
 since the currently-expected clump-forward shock impact is not observed. 
Recent studies nonetheless identified both Tycho's SN and SN 1006 as normal Type Ia supernovae.

The global approximations of Type Ia models with exponential models 
may not be sufficiently accurate enough close to the contact discontinuity, particularly for the PDD1c model, 
where the ejecta appears to diverge from
the exponential model at low velocities. 
 If the ejecta density profile deviates from the exponential or power-law models,
   a higher clump speed and an earlier clump-remnant impact is expected for a flatter ejecta substrate, 
 and vice versa.
However, significant deviations from the exponential model are not anticipated  
for the W7 and DD200c models.
 The clump-remnant impact is estimated to occur within $\sim 2000$ yrs under  
  all plausible explosion conditions of SNe Ia. 
As clumps are expected to be present only in a thin shell at the boundary 
of the Ni bubble, not throughout the bulk of the ejecta,
 the small clump size and a flat bubble density contrast should not induce
  vigorous mixing in  
  the ensuing clump/bubble-remnant interactions,
 which such characteristic distinguishes the remnants of SNe Ia from those of the core-collapse SNe.

The author would like to thank Roger Chevalier, Dietrich Baade and Vikram Dwarkadas 
for useful discussions and correspondence,
and the anonymous referee for valuable comments on the manuscript. 
This work was supported by the National Science Council of the Republic of China, Taiwan, and 
the National Center of High-Performance Computing of Taiwan.  
Ted Knoy is appreciated for his editorial assistance.




\plotone{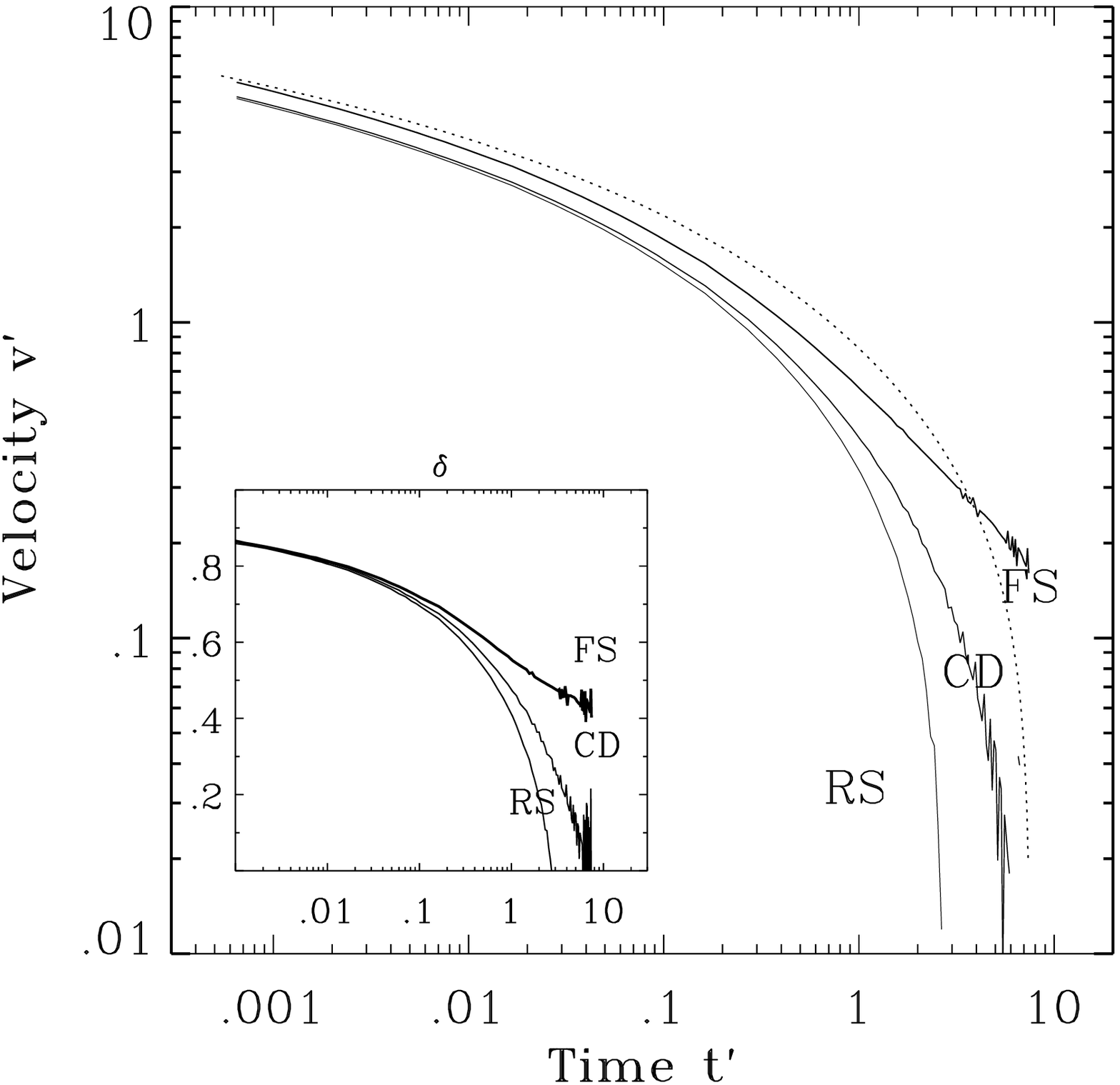}
\figcaption[f1.eps]
{Evolution of the characteristic velocities of an SNR that arises from 
interaction of exponentially-declining ejecta 
with a uniform ambient medium, as described in DC98 and WC01.
Dotted line: Flow velocity of ejecta immediately below reverse shock (RS).
Solid lines: Pattern velocities and deceleration parameters
$\delta=dlnr/dlnt$ (inset) of forward shock (FS),
contact discontinuity (CD) and reverse shock.
\label{fig.1}}

\vspace{10mm}
\begin{figure}[!hbtp]
\centerline{\includegraphics{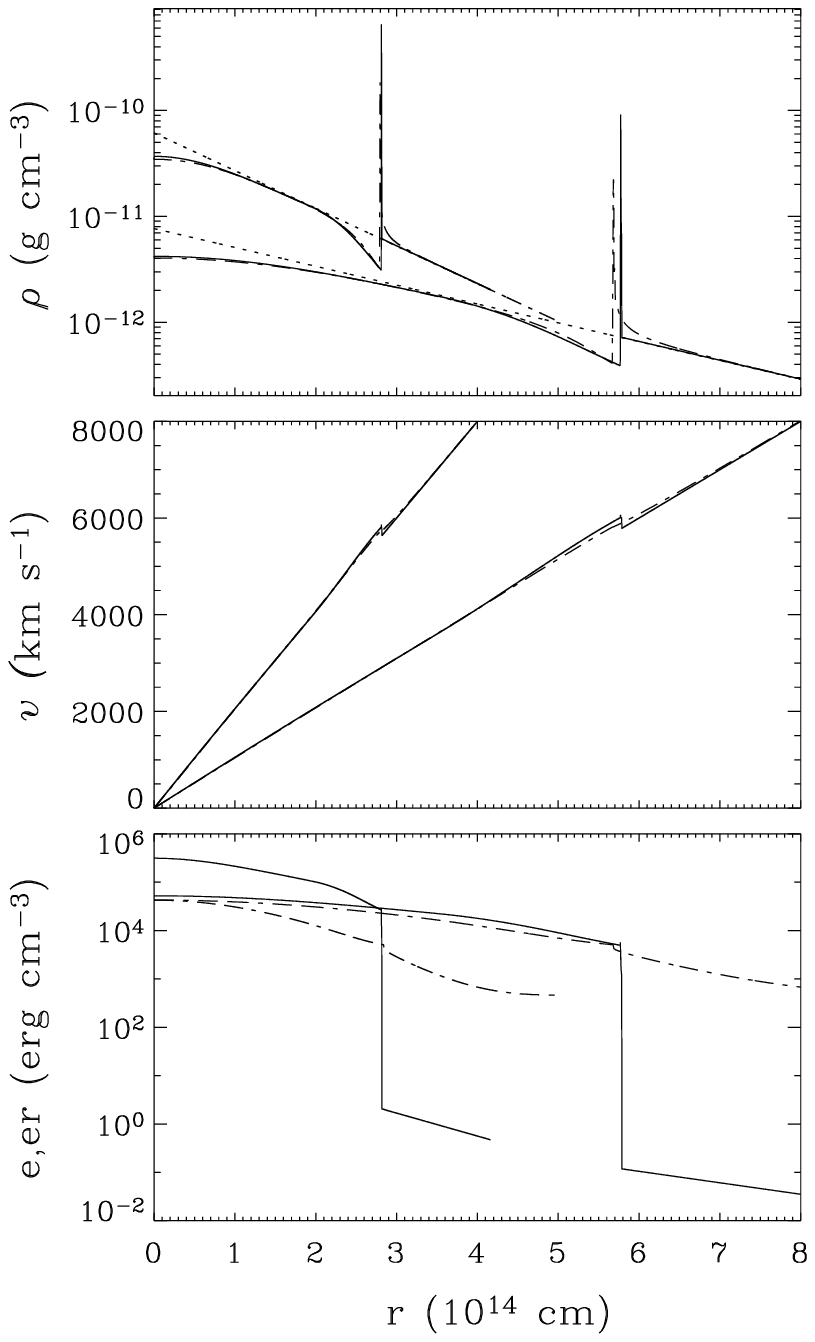}}
 \caption{
Distributions of material density, velocity and thermal or radiative energy density 
of Ni bubble in standard exponential model
 at $5\times 10^{5}$ and $10^{6} \seco$. 
 Solid and dash-dotted lines represent adiabatic and radiative diffusive cases, respectively. 
 Also plotted as dotted line is unaltered exponential density distribution.
 HD runs exploit a grid of 16000 uniform zones, resolving the shell into $\ga 100$ zones,
 while RHD runs exploit a grid of 8000 uniform zones. 
 \label{fig.2}}
\end{figure}

\plotone{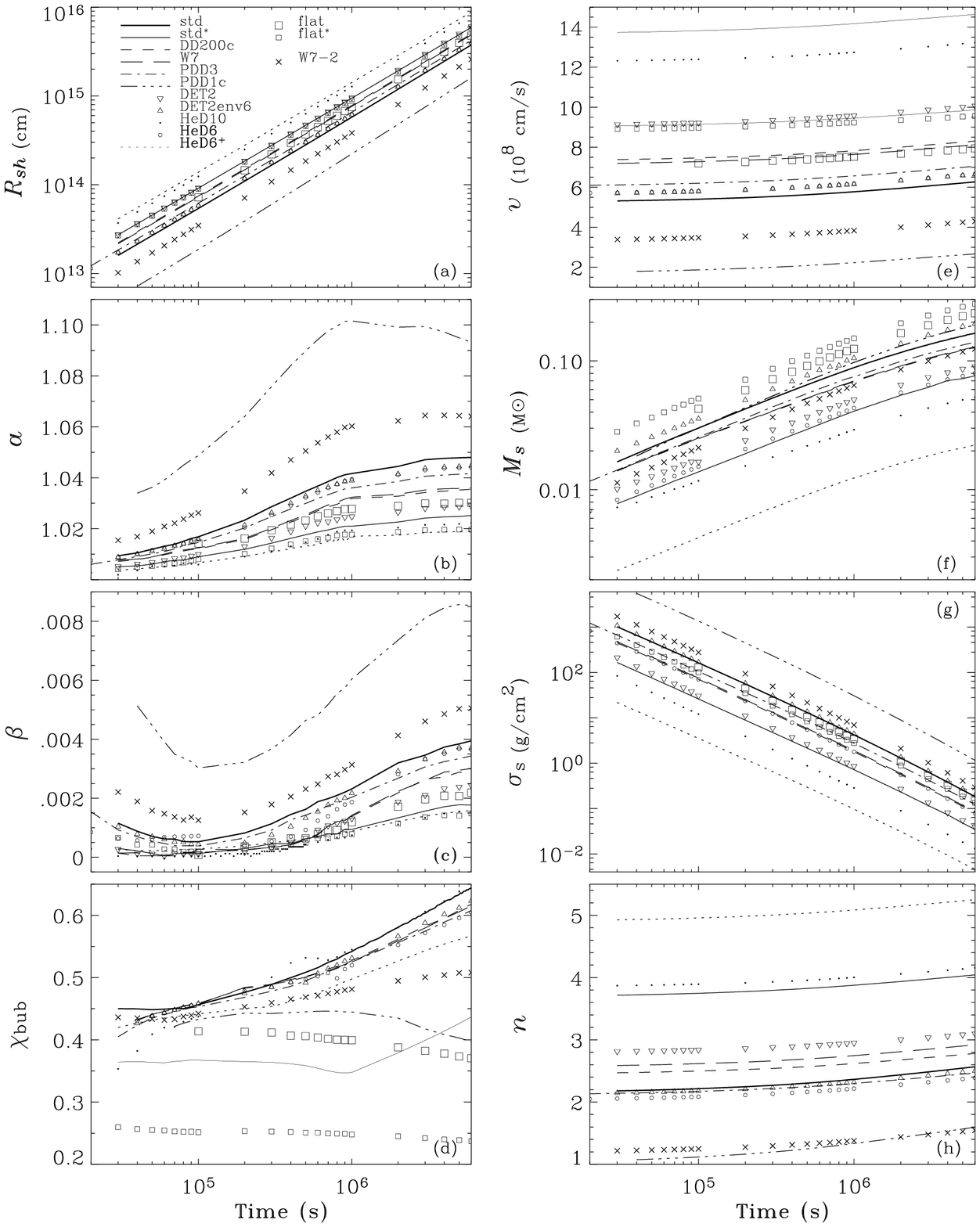}

\figcaption[f3.eps]
{
Evolution of properties of Ni bubble shell in various models. 
 The grids use 16000 uniform zones for all runs.
(a) 
Radius $R_{sh}$ evaluated at shock front.
(b)
Expansion rate $a=dlnR_{sh}/dlnt$.
(c)
Thickness ratio $\beta$.
(d)
 Density contrast (middle) of bubble interior near contact discontinuity, 
 relative to the ambient ejecta substrate. 
 In this plot, the PDD3, W7 and DD200c models appear to yield matching results.
(e)
  Flow velocity $v_{flow,sh}=R_{sh}/t$ evaluated at densest point of shell (CD). 
(f)
Swept-up mass $M_s$. 
(g)
Surface density $\sigma_s \equiv M_s / 4 \pi R_{sh}^2$.
(h)
Power index $n$ of the density of the ejecta substrate at contact discontinuity.
\label{fig.3}}

\vspace{10mm}
\begin{figure}[!hbtp]
\centerline{\includegraphics{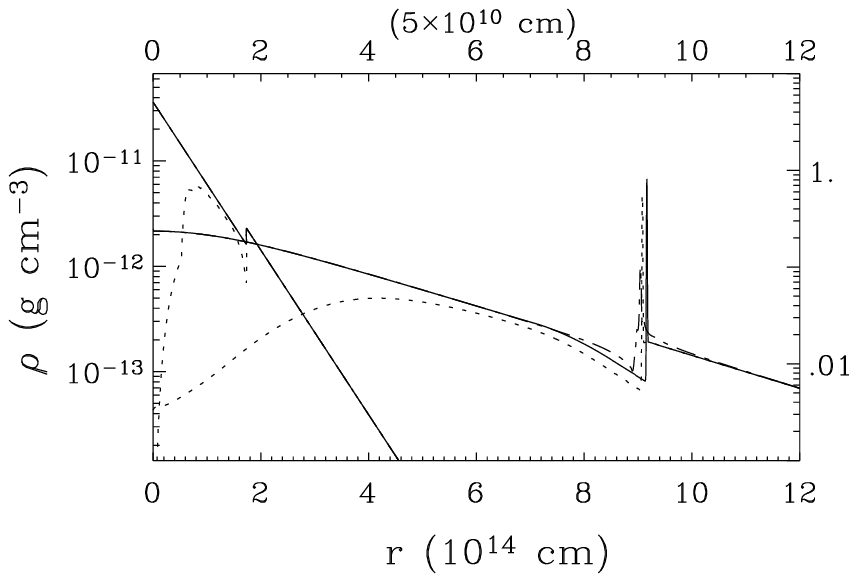}}
\caption{
Ni bubble density profiles at $10^2 \seco$ (beginning of simulations, top and right axes) and $10^6 \seco$ (bottom and left axes) for the exponential $\omega=0.7$ W7 model (solid line),
 RHD counterpart (dashed-dotted line),
and modified exponential model using initial $\Ni$ mass fraction given by Nomoto {\it et al.} (dotted line). 
The RHD model uses grid with 2000 uniform zones and HD models use grid with 4000 uniform zones.
\label{fig.4}}
\end{figure}

\vspace{10mm}
\begin{figure}[!hbtp]
\centerline{\includegraphics{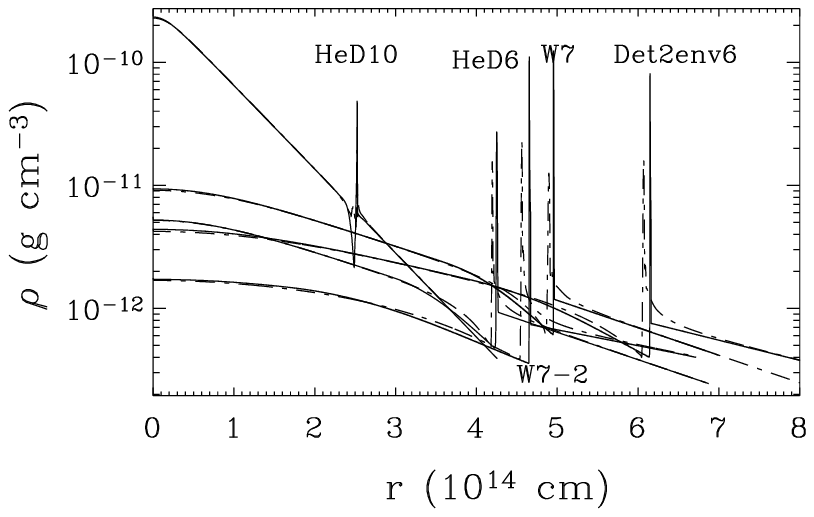}}
\caption{
Ni bubble density profiles in RHD models at their estimated frozen-in ages.
Models from left to right are, respectively, HeD10 ($2\times 10^5 \seco$), HeD6 ($7\times 10^5 \seco$), W7-2 ($1\times 10^6 \seco$),
W7 ($1\times 10^6 \seco$) and DET2env6 ($1\times 10^6 \seco$).
Solid lines represent adiabatic cases and dash-dotted lines represent radiative cases.
Model HeD10 uses grid with 2000 uniform zones while all other runs use grid with 4000 uniform zones.
\label{fig.5}}
\end{figure}

\vspace{10mm}
\begin{figure}[!hbtp]
\centerline{\includegraphics{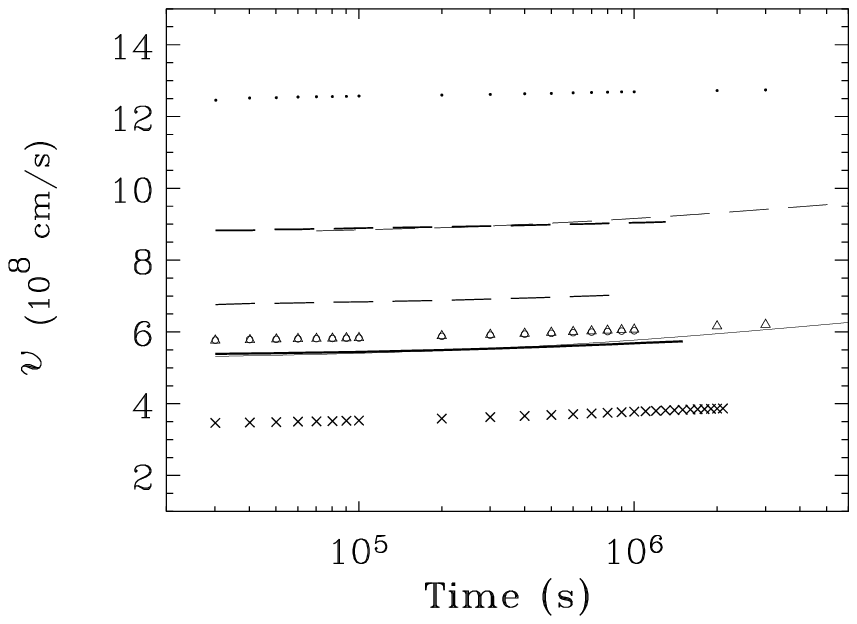}}
\caption{
Top: Flow velocities $v_{flow,sh}=R_{sh}/t$ evaluated at CD
 in radiative models.
 Models from top to bottom are the HeD10, $\omega=0.7$ W7 (both HD and RHD cases), W7, 
 DET2env6, HeD6, std (HD and RHD cases) and W7-2, respectively.
%
\label{fig.6}}
\end{figure}

\plotone{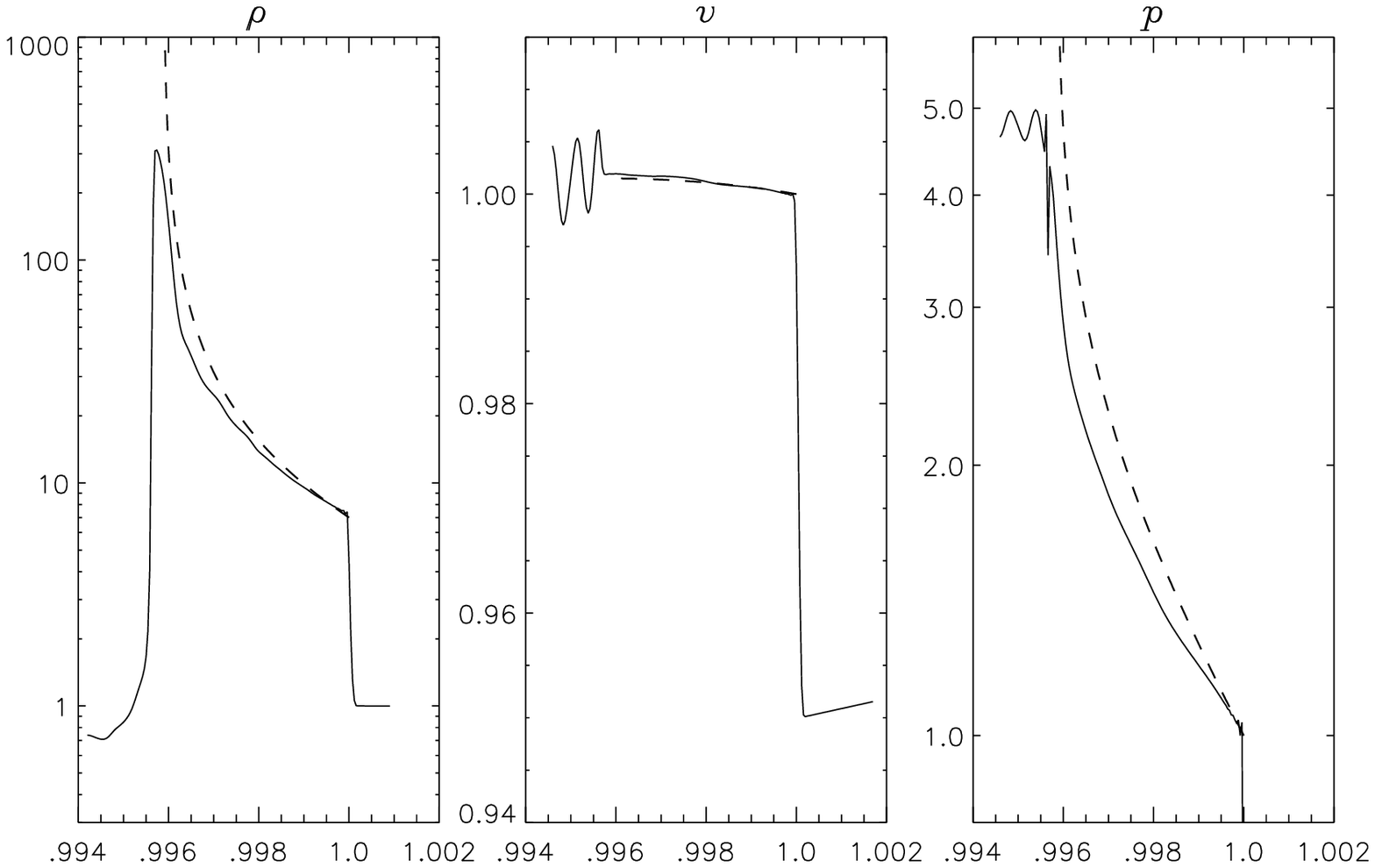}

\figcaption[f7.eps]
{Hydrodynamical solutions of Ni bubble shell in standard model
at $2.0 \times 10^{6} \seco$ 
overplotted on self-similar solution for $n=2.448$ and $l=0.8958$ (corresponding to $a=1.041$).
These self-similar solutions were obtained by integrating 
the self-similar differential equations using the fourth-order Runge-Kutta method.
The shell is resolved into $\sim 140$ grid zones 
in the HD solutions and 200 zones in the self-similar solutions. 
The HD solutions exhibit large oscillations in the velocity and pressure distributions
behind the CD, which are caused by the use of the shock-capturing scheme 
which aims to resolve an abrupt density change 
in a molecular length scale into a few grid zones,
whose numerical origin is the same as that of the postshock oscillations.  
Traditionally, 
 weaker shocks and higher numerical resolution are thought to produce more severe postshock oscillations, 
whose presence nevertheless depends on the problem; 
satisfactory design criteria 
that ensure that the captured shocks are simultaneously narrow and free from oscillations 
do not exist.
For a discussion of postshock oscillations, see Arora and Roe (1997).
\label{fig.7}}

\vspace{10mm}
\begin{figure}[!hbtp]
\centerline{\includegraphics{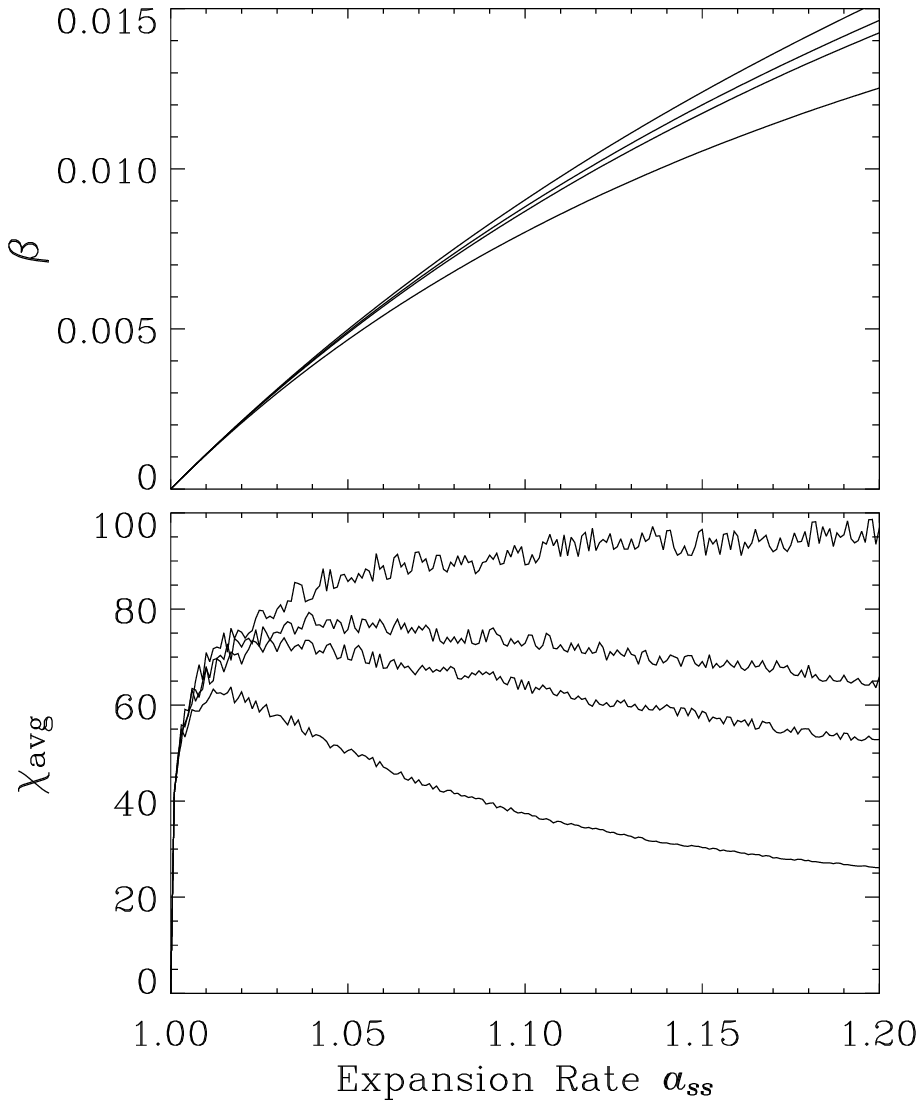}}
\caption{
 Variations in self-similar thickness ratio 
 and mean density contrast 
 with expansion rates for four density power indices.
The lines from top to bottom refer to cases of $n=3$, $n=2.4$, $n=2$ and $n=0$. 
 The solutions were obtained using a fixed integration step size of $6.6\times 10^{-7}$. (Radius of shock front is one).
\label{fig.8}}
\end{figure}

\vspace{10mm}
\begin{figure}[!hbtp]
\centerline{\includegraphics{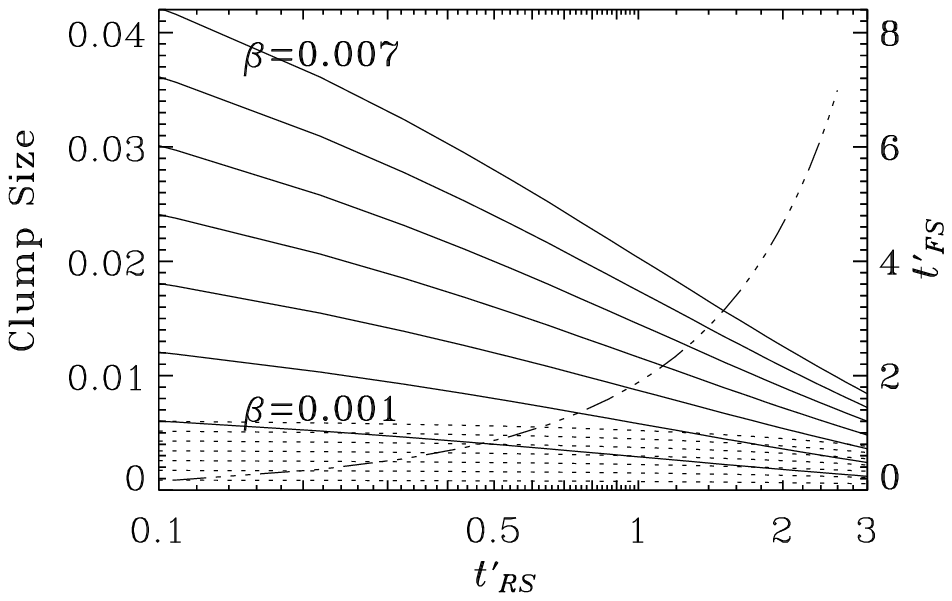}}
\caption{
Initial clump size (solid and dotted lines) and forward shock impact time (triple-dot-dashed line)
as functions of reverse shock impact time.  
Solid lines: Size as fraction of intershock width.
Dotted lines: Size as fraction of forward shock radius. 
Lines from top to bottom refer to cases $\beta=0.007$, 0.006, 0.005,
      0.004, 0.003, 0.002 and 0.001, respectively.
\label{fig.9}}
\end{figure}

\vspace{10mm}
\begin{figure}[!hbtp]
\centerline{\includegraphics{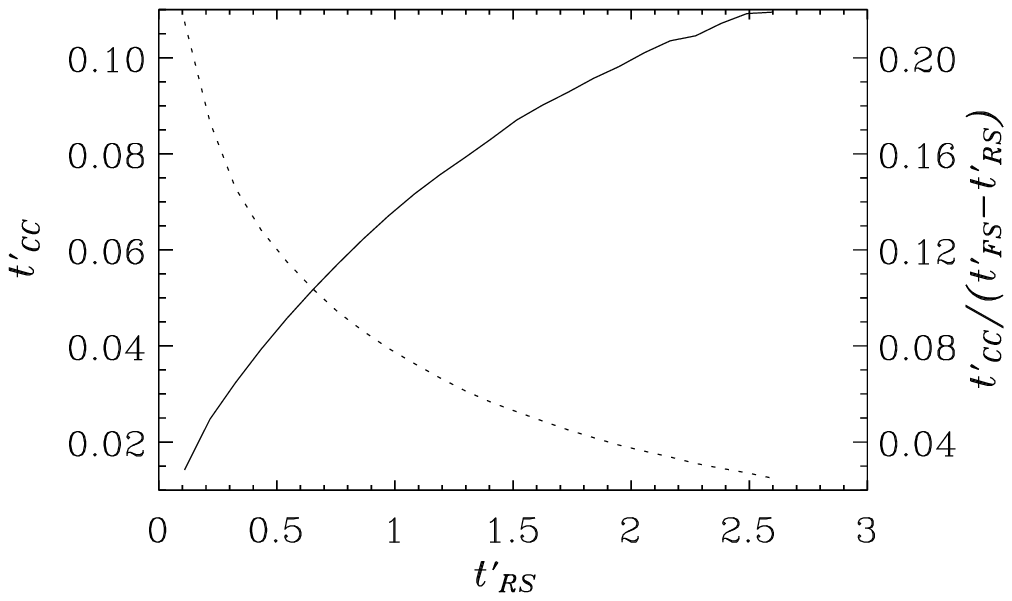}}
\caption{
Cloud crushing time (solid line) and
 ratio of cloud-crushing time to 
 intershock-crossing time (dotted line) 
as functions of reverse shock impact time,  
assuming 
 $t_{cc}= \beta r_{RS} {\chi}^{1/2} /(v_{f}-v_{RS})$,
 $\beta=0.04$ and $\chi=100$ for the clump.
\label{fig.10}}
\end{figure}

%
%

\clearpage

  \include{tab1}

\end{document}

%% file: tab1.tex
 \begin{table}[h]
\small{
 \begin{center}
 \caption{SN Ia EXPONENTIAL MODELS \label{tabmodel}}
 \begin{tabular}{llllrcrrrcc}
 \tableline
 \tableline
 Model             \tablenotemark{1} &
 $M$               \tablenotemark{2} &
 $E_{51}$          \tablenotemark{3} &
 $M_{Ni}$          \tablenotemark{4} &
 $A$               \tablenotemark{5} &
 $v_{e}$           \tablenotemark{6} &
 $v_{f}$           \tablenotemark{7} &
 $v'_{f}$          \tablenotemark{8} & 
 $t'_{RS}$          \tablenotemark{9} & 
 $t_{RS}(1/0.5)$    \tablenotemark{10} &  
 $t'$              \tablenotemark{11} \cr 
 \tableline
       std &  1.4  & 1.0  & 0.50  &  7.7 &  2439 &  5815 & 0.69 & 1.34 &  332/ 419 & 1.75/1.39
\\
      std* &  -  & -  & -  &  - &  - &  9300 & 1.10 & 0.59 &  146/ 184 & -
\\
    DD200c &  1.4  & 1.5  & 0.613 &  4.2 &  2988 &  7745 & 0.75 & 1.18 &  238/ 301 & 2.14/1.70 
\\
        W7 &  1.4  & 1.3  & 0.59  &  5.2 &  2781 &  7558 & 0.78 & 1.09 &  237/299 & 1.99/1.58
\\
        W7($\omega=0.7$) &  -  & -  & -  &  - &  - &  9030 & 0.94 & 0.80 &  174/219 & -/-
\\
      flat &  1.4  & 1.0  & 0.50  &   NA  &  NA  &  7528 & 0.89 & 0.88 &  217/ 274 & 1.75/1.39
\\
     flat* &  -  & -  & -  &  - &  - &  9219 & 1.09 & 0.60 &  148/ 187 & - 
\\
      PDD3 &  1.4  & 1.37 & 0.49  &  4.8 &  2855 &  6543 & 0.66 & 1.42 &  301/ 379 & 2.04/1.62
\\
     PDD1c &  1.4  & 0.47 & 0.10  & 23.8 &  1672 &  2526 & 0.44 & 2.43 &  877/1106 & 1.20/0.95
\\
      DET2 &  1.2  & 1.52 & 0.63  &  2.8 &  3248 &  9413 & 0.84 & 0.98 &  173/ 219 & 2.45/1.94
\\
  DET2env6 &  1.2+0.6 & 1.52 & 0.63  &  7.7 &  2652 &  6191 & 0.67 & 1.38 &  343/ 432 & 1.75/1.39
\\
     HeD10 &  0.8+0.22 & 1.24 & 0.75(0.1) &  2.5 &  3182 & 12474 & 1.13 & 0.56 &   95/ 119 & 2.53/2.01
\\
      HeD6 &  0.6+0.172 & 0.72 & 0.252(0.08) &  2.8 &  2787 &  6093 & 0.63 & 1.52 &  270/ 340 & 2.43/1.93
\\
     HeD6+ &  - & - & 0.08  &  - &  - & 13833 & 1.43 & 0.33 &   58/  73 & -
\\
         W7-2 &  1.4  & 1.3  & 0.16  &  5.2 &  2781 &  3948 & 0.41 & 2.59 &  564/ 710 & 1.99/1.58
 \\
        W7-2* &  -  & -  & -  &  - &  - &  4992 & 0.52 & 1.98 &  431/ 543 & - 
 \\
 \end{tabular}
 \end{center}
  \tablenotetext{ }{Quantities given in columns are as follows:
    (1) name of model; names with * are cases with $\omega=0.5$  
    (2) supernova mass ($M_\odot$)   
    (3) supernova kinetic energy ($10^{51} \rm ergs$)  
    (4) initial mass of $\Ni$ ($M_\odot$) 
    (5) constant of the exponential model ($10^6$ cgs units)  
    (6) velocity scale height ($\rm km~s^{-1}$)
    (7) frozen-in velocity ($\rm km~s^{-1}$)
    (8) normalized frozen-in velocity 
    (9) normalized age of the reverse shock impact 
    (10) age (yr) of the reverse shock impact for an ambient density of $n_{am}=1$ and 0.5, respectively  
    (11) normalized age of Tycho's remnant for an ambient density of $n_{am}=1$ and 0.5, respectively. } 
} 
 \end{table}